\documentclass{aa}

\usepackage{graphicx}
\usepackage{txfonts}
\usepackage{natbib,twoopt}
\bibpunct{(}{)}{;}{a}{}{,}    

\usepackage{multirow}
\usepackage[dvipsnames]{xcolor}
\usepackage{makecell}
\usepackage{hyperref} 
\hypersetup{
  colorlinks = true, 
  urlcolor   = blue, 
  linkcolor  = blue, 
  citecolor  = blue 
}
\usepackage{soul}

\makeatletter
\newcommand\newtag[2]{{#1}\def\@currentlabel{\textsuperscript{(#1)}}\label{#2}}
\makeatother

\newcommand{\MgI}{\ion{Mg}{I}\xspace}
\newcommand{\MgII}{\ion{Mg}{II}\xspace}
\newcommand{\MgIII}{\ion{Mg}{III}\xspace}

\makeatletter
\renewcommand*\aa@pageof{, page \thepage{} of \pageref*{LastPage}}
\makeatother

\begin{document}

\title{Modeling the \MgI from the NUV to MIR}
\subtitle{II. Testing stellar models}

\author{
    J. I. Peralta\inst{1,2}
    \and M. C. Vieytes \inst{1,2}
    \and A. M. P. Mendez\inst{1}
    \and D. M. Mitnik \inst{1,3}
}

\institute{Instituto de Astronomía y Física del Espacio, CONICET--Universidad de Buenos Aires, Argentina\\
        \email{jperalta@iafe.uba.ar; mariela@iafe.uba.ar}
        \and Departamento de Ciencia y Tecnología, UNTREF, Argentina
        \and Departamento de Física, Universidad de Buenos Aires, Argentina
}

\date{}

\abstract
{
Reliable atomic data are mandatory ingredients to obtain a realistic semiempirical model of any stellar atmosphere. Due to their importance, we further improved our recently published \MgI atomic model. 
}
{
We tested the new atomic model using atmospheric models of stars of different spectral types: the Sun (dG2), HD22049 (dK2, Epsilon Eridani), GJ 832 (dM2), and GJ 581 (dM3). 
}
{
Significant improvements have been included in the atomic model, mainly to the electron impact excitation ($\Upsilon_{ij}$) values. We used new Breit-Pauli distorted-wave (DW) multiconfiguration calculations, which proved to be relevant for many transitions in the mid-infrared (MIR) range.
The new atomic model of \MgI includes the following: \textit{i}) recomputed $(\Upsilon_{ij})$ data through the DW method, including the superlevels. \textit{ii}) For the nonlocal thermodynamic equilibrium (NLTE) population calculations, 5676 theoretical transitions were added (3001 term-to-term). \textit{iii}) All of these improvements were studied in the Sun and the stars listed above. Comparisons for the distribution of magnesium among the first ionization states and the formation of molecules, as well as for the population of the different energy levels and atmospheric heights, were carried out. Several lines, representative of the spectral ranges, were selected to analyze the changes that were produced. In particular, we exemplify these results with the problematic line 2853.0 {\AA}, a transition between the third level and the ground state.
}
{
The magnesium distribution between ionization states for stars with different effective temperatures was compared. For the Sun and Epsilon Eridani, \MgII predominates with more than 95\%, while for GJ 832 and GJ 581, \MgI represents more than 72\% of the population. Moreover, in the latter stars, the amount of magnesium forming molecules in their atmosphere is at least two orders of magnitude higher. Regarding the NLTE population, a noticeable lower variability in the departure coefficients was found, indicating a better population coupling for the new model. 
Comparing the synthetic spectrum calculated with the older and new \MgI atomic model, these results show minimal differences in the visible range but they are stronger in the infrared (IR) for all of the stars. This aspect should be considered when using lines from this region as indicators. Nevertheless, some changes in the spectral type were found, also emphasizing the need to test the atomic models in different atmospheric conditions. 
The most noticeable changes occurred in the far-ultraviolet (FUV) and near-ultraviolet (NUV), obtaining a higher flux for the new atomic model regardless of the spectral type. The new model did not prevent the formation of the core emission in the synthetic NUV line 2853.0 {\AA}. However, by including other observations, we could note that the emission indeed exists, although with a much lower intensity. Further tests have shown that to reduce the emission, the population of its upper level (3s3p $^1P$ ) should be reduced by a factor of about 100.
}
{}
 
\keywords{Atomic data -- Stars: late-type -– line: formation -– line: profiles}

\maketitle

\section{Introduction} \label{sec:intro}

\begin{table*}[t]
\caption{Summary of atomic models.} 
\label{table:models}
\centering
\begin{tabular}{cccll}
\hline\hline
Model & \MgI levs & \MgII levs & \multicolumn{1}{c}{$\Upsilon_{ij}$ methods (\MgI)} & \multicolumn{1}{c}{Details}\\ 
 \hline
 base  & 26 & 14 & SEA\&VRM & Original model.\textsuperscript{(1)} \\ \hline
 c & 85 & 47 & \begin{tabular}{@{}l@{}}
                CCC (< lev 26) + DW (levs 26 to 54) +\\
                SEA\&VRM (levs 55 to 85)\end{tabular} & 
                \begin{tabular}{@{}l@{}} 
                \textit{gf}, broadening and photoionization data updated.\textsuperscript{(2)}\\
                \end{tabular}\\\hline
 d & 85 & 47 & CCC (< lev 26) + DW (> 25) & 
                \begin{tabular}{@{}l@{}}
                \textit{gf}, broadening and photoionization data updated.\\
                 $A_{ij}$ for NLTE from \textit{AUTOSTRUCTURE}
                 \end{tabular}\\\hline
\hline
\end{tabular}
\tablefoot{In each star, the atomic model was changed but the atmospheric structure remained unchanged.}
\tablebib{
(1) \citet{fontenla:2015}; (2) \cite{peralta:2022}
}
\end{table*}

In the last decades, the discovery of many exoplanets orbiting in the habitable zone of late-type stars has shown the need for reliable calculations of the spectral energy distribution (SED) of those stars, from IR to X-rays. Obtaining these values is a task suitable for well-established stellar atmospheric models.

Stellar models are essential for obtaining information inaccessible to direct observation, such as from the extreme-ultraviolet (EUV) radiation or Ly$\alpha$ emission. They also allow for the indirect estimation of the characteristics of a star, including luminosity, magnetic activity, and metallicity. Using a synthetic spectrum, it is possible to calculate the amount of nonthermal energy that must be delivered to the atmosphere to reproduce the observed spectrum, thus constraining the possible physical processes involved.

When improving an atomic model, testing it in the solar atmosphere is the best starting point. The possibility of observing with spatial resolution allows us to improve the accuracy of the model and even the capacity to model the different structures present in the solar atmosphere separately \citep{fontenla:2006,fontenla:2007,fontenla:2009}. In \cite{peralta:2022} (Paper I, hereafter), we show the importance of having reliable atomic data, which allows the atmospheric model to achieve a realistic behavior when calculating the atomic populations in situations out of local thermodynamic equilibrium (LTE), and thus produce spectral lines according to the observations. 

The calculation of an atmospheric model depends on relevant atomic parameters at the quantum scale to describe the physical processes that take place and on the thermodynamic parameters of the plasma that comprises it. Since the same atomic model is used under different conditions, it is essential to test its validity in plasma with different thermodynamic parameters, which is the case of stellar atmospheres of different spectral types. Several authors performed this type of study with neutral magnesium for different purposes and methods, as we describe in Paper I (e.g., \cite{alexeeva:2018, barklem:2017, osorio:2015}, etc.).
More recently, and motivated by the controversy over the abundance of the solar photospheric oxygen, \cite{bergemann:2021} improved the calculation by building a more reliable oxygen atomic model. They followed a similar atomic-parameter approach to the one included in our work in Paper I for \MgI; the authors calculated the excitation rates by electron collisions using the Breit-Pauli distorted wave (DW) method through the \textsc{autostructure} code \citep{badnell:2011}. However, none of the aforementioned works reproduce spectral lines of \MgI in the UV, mid-infrared (MIR), and far-infrared (FIR), except for particular lines at micrometer wavelengths. dM stars are also not included in their samples.

Since the preferred targets for planet-hunting are dM stars, and this spectral type is bright in the IR range, it is important to have a reliable estimate of the radiation in this range to characterize these objects. The successful launch and deployment of the James Webb Space Telescope (JWST), which operates in the visible and MIR, emphasizes this need. On the other hand, in addition to the IR region, photochemistry and mass loss in planetary atmospheres are caused by UV radiation. For this reason, its correct description is essential for characterizing an exoplanetary atmosphere \citep{france:2016}.

One of the most prominent features of \MgI in the near-ultraviolet (NUV) range is the 285.3 nm ($3s^2\,^1S_0 - 3p\,^1P_1$) line profile, which formed in the chromosphere of late-type stars. \cite{fontenla:2016}, and then \cite{tilipman:2021} built atmospheric models of the dM star GJ 832 from the photosphere to the corona using solar-stellar radiation physical modeling (SSRPM) \citep{fontenla:2016}. SSRPM is a variant of solar radiation physical modeling (SRPM) version 2 (presented in \cite{fontenla:2015}, and detailed in Paper I), which extends atmospheric modeling to stars of different spectral types. The atmospheric models produced in both papers show a good overall fit of the observed spectrum but an incorrect fit of the 285.3 nm line. According to the authors, this behavior occurs due to uncertainties in the \MgI atomic data.

For this work, we have extended the study carried out in Paper I in two ways: on the one hand, we generated an improved version of the atomic model of \MgI presented in that paper; and on the other hand, we tested this new atomic model on the atmospheric models of the Sun, and on three stars of spectral types cooler than the Sun's. The new atomic model shares the same features as the ``1401c'' (``c,'' hereafter) of Paper I, but we improved the radiative and electron impact excitation data of \MgI for the higher levels. Both models have identical data for the effective collision strengths ($\Upsilon_{ij}$) up to and including level 54 ($3s7i\,^1I$, 59\,430.52 $cm^{-1}$). That is, they include data computed by \cite{barklem:2017} via the convergent close-coupling (CCC) method for the lowest 25 levels and the DW calculations of Paper I for transitions between levels 26 ($3s5g\,^1G$, 57\,262.76 $cm^{-1}$) and 54. However, in our new atomic model, the difference lies in the data for transitions from (and to) levels with energies higher and equal to superlevel 55 (59\,649.15 $cm^{-1}$). For these levels, we included electron impact excitation values obtained by using the DW approximation implemented in the \textsc{autostructure} code. In this way, the 85-level model (with superlevels from level 55 onward) no longer uses the, already traditional, semiempirical methods of \cite{seaton:1962} and \cite{vanRegemorter:1962} for the calculation of $\Upsilon_{ij}$ - employed for nonlocal thermodynamic equilibrium (NLTE) population computations - in transitions that are extremely important for the formation of lines up to the MIR.
Regarding the atmospheric models selected, we used the solar model presented in Paper I by \cite{fontenla:2015} and three other models:
HD 22049 dK2V (``Epsilon Eridani,'' hereafter) by \cite{vieytes:2021}, 
GJ 832 dM2 ($T_\text{eff}=3590$~K and $R/R_{\odot}=0.5$), and GJ 581 dM3 ($T_\text{eff}=3498$~K and $R/R_{\odot}=0.3$) by \cite{tilipman:2021}.

This paper is structured as follows: The code used in our NLTE and spectra calculations and the initial atomic model are detailed in Section \ref{sec:NLTE_calculations}. Section \ref{sec:new_model} shows the different \MgI atomic models built. A general description of the atmospheric models for the Sun and the three abovementioned stars is given in Section~\ref{sec:atmos_models}. The observations used for comparing our synthetic spectra are described in Section \ref{sec:observations}. Our results and discussion are presented in Section \ref{sec:results}, while our final remarks and conclusions are provided in Section \ref{sec:conclusions}.

\begin{figure}[t]
\centering
\resizebox{\hsize}{!}{\includegraphics{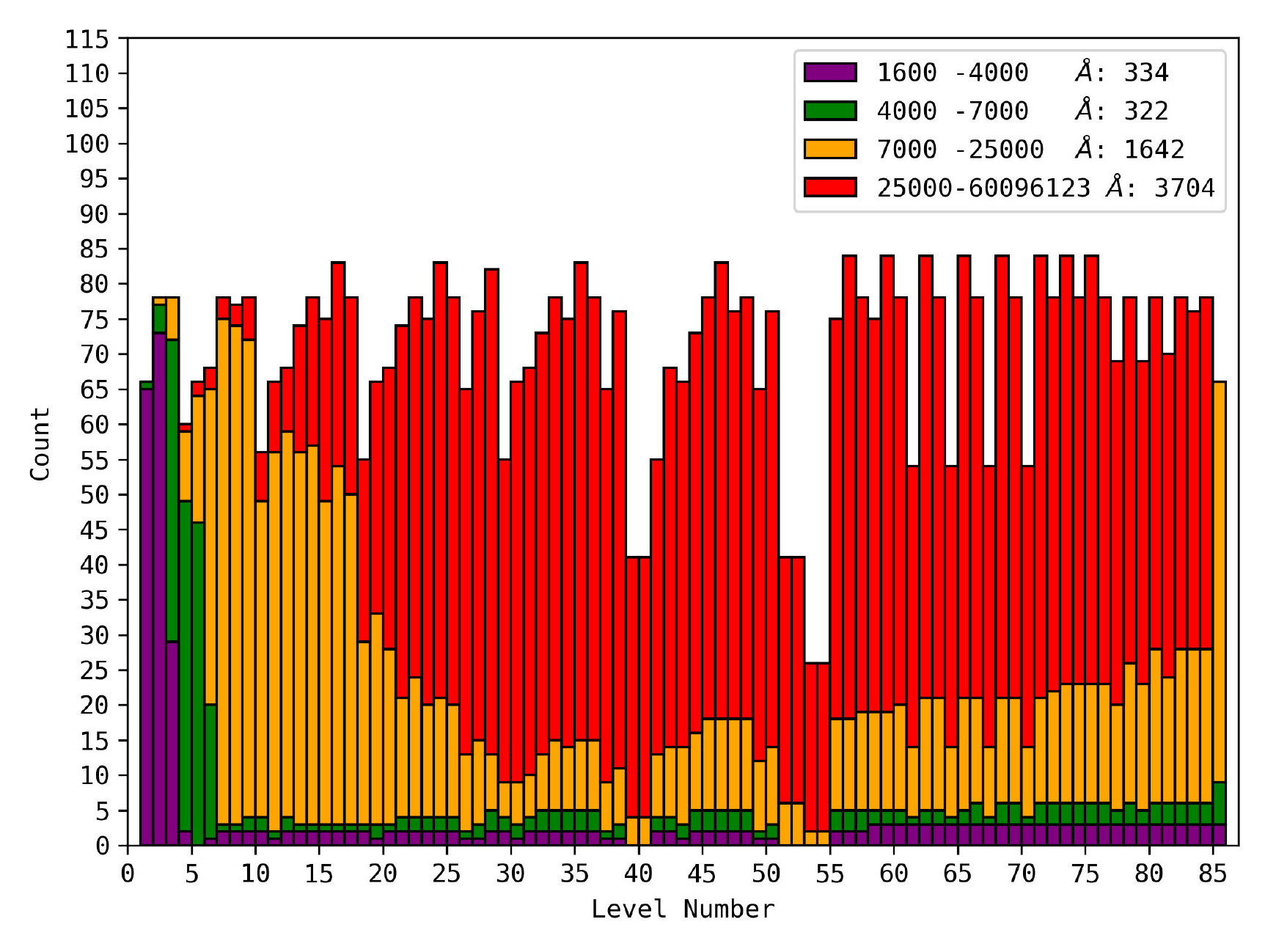}}
\caption{Stacked histogram of the levels used by the 3001 radiative term-term transitions in the NLTE population calculation of model d. The purple bars represent the levels used at wavelengths in the FUV and NUV, the green bars in the visible range, the orange bars in the NIR, and the red bars from the MIR to the FIR.}
\label{fig:histogram}
\end{figure}

\section{NLTE population and spectrum calculations} \label{sec:NLTE_calculations}

\begin{figure*}[ht]
\centering
\includegraphics[width=.5\textwidth]{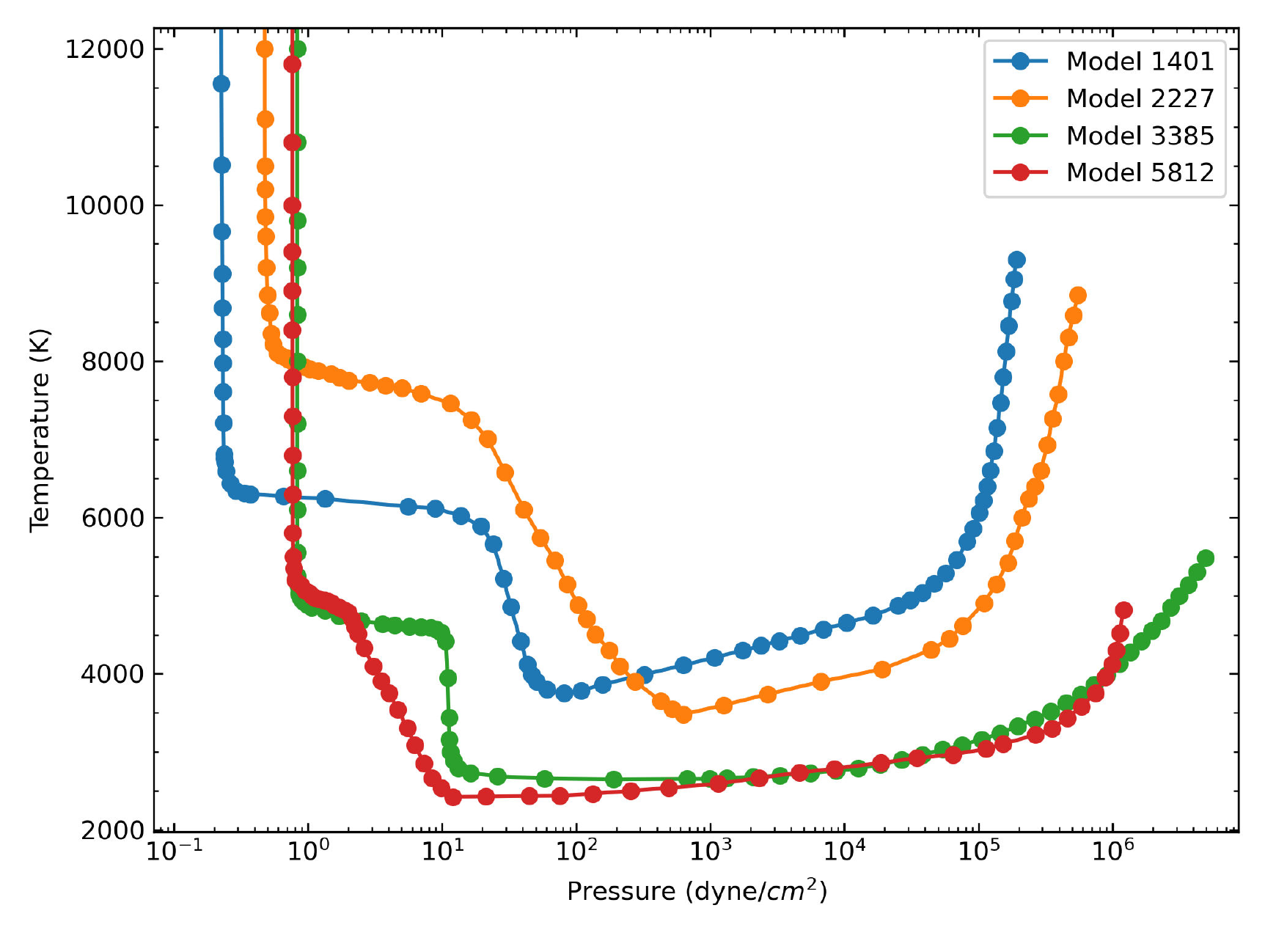}
\includegraphics[width=.5\textwidth]{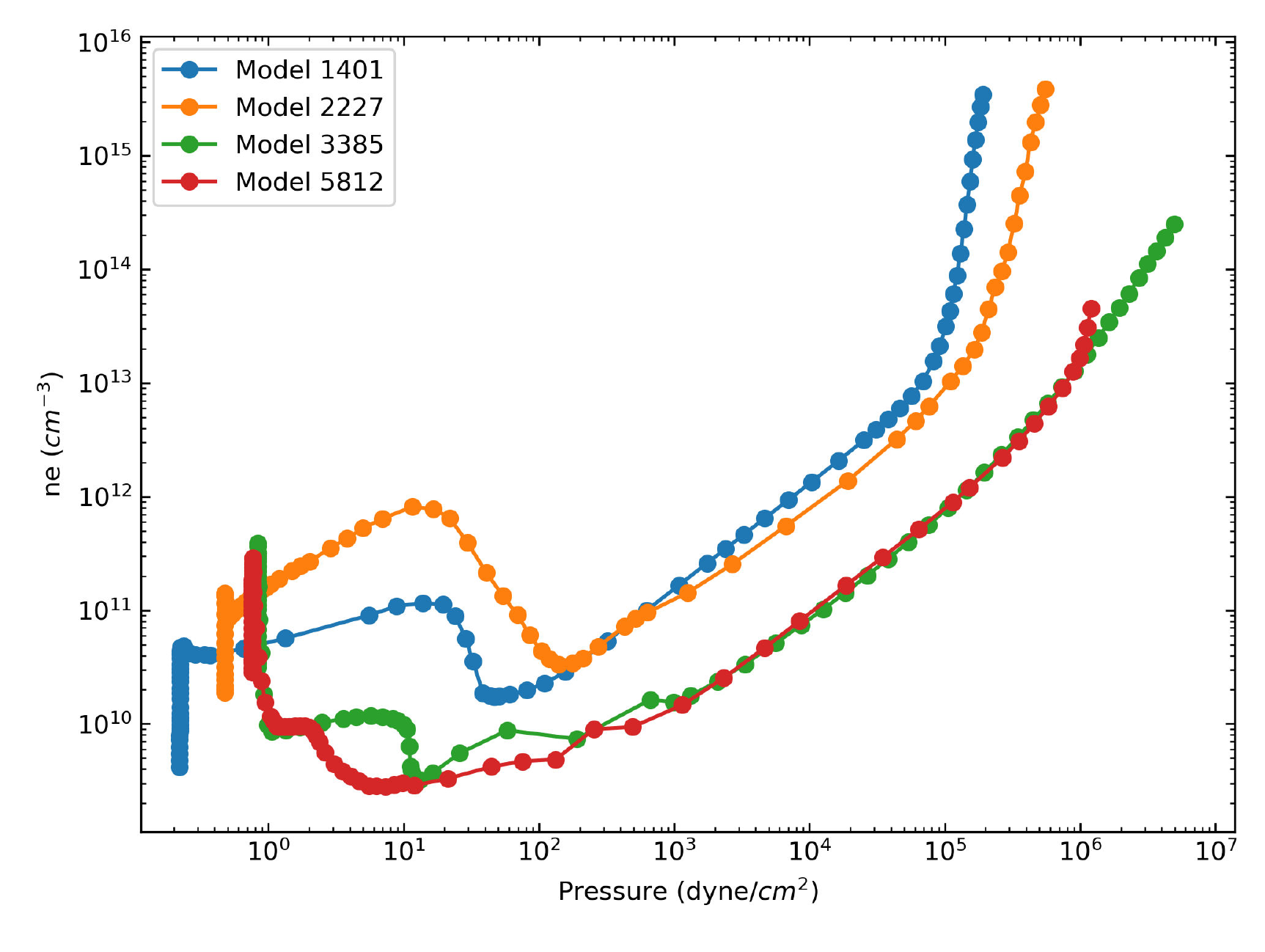}\vspace{0cm}
\caption{Temperature (left panel) and electron density (right panel) as a function of pressure, for atmospheric base models 1401, 2227, 3385, and 5812, corresponding to the Sun (in blue), Epsilon Eridani (in orange), GJ 832 (in green), and GJ 581 (in red), respectively.}
\label{fig:T_ne_vs_press}
\end{figure*}



The general procedure for calculating the populations in NLTE and the emergent spectrum is generated similarly as we described in Paper I. In this case, the Solar Stellar Radiation Physical Modeling (SSRPM) version 2 \citep{fontenla:2016} was used. SSRPM is a variant of the SRPM used in Paper I \citep{fontenla:2015}, which extends the atmospheric modeling to stars cooler than the Sun. As well as the SRPM code, this library assumes hydrostatic equilibrium and solves the statistical equilibrium and radiative transport equations in a self-consistent way for an atmosphere with plane-parallel or spherical symmetry. For the NLTE atomic populations calculation, in an optically thick atmosphere and including partial redistribution (PRD), the code contains 52 neutral and low ionization state atomic species (generally up to $Z^{2+}$), H, H$^-$, and H$_2$. In addition, it allows the calculation of 198 highly ionized species using the optically thin atmosphere approximation. The SSRPM adds the calculation of molecule formation in LTE, which includes molecular sequestration of elements. For spectral lines, 435\,986 transitions are included, along with more than 2\,000\,000 molecular lines (with data from \cite{plez:1998}) of the 20 most abundant and important diatomic molecules for dM stars (e.g., TiO). In other words, it allows the construction of atmospheric models not only for the different observed characteristics on the solar disk but also for stars of different spectral types. In \cite{fontenla:2016}, the model for the dM star GJ 832 was presented, while the model for Epsilon Eridani was built in \cite{vieytes:2021}. In \cite{tilipman:2021}, the atmospheric model of GJ 832 is improved, and a new model for the dM star GJ 581 is presented. The model of Epsilon Eridani and the latest versions of GJ 832 and GJ 581 will be used in this work.

\section{Convergent close coupling \& distorted wave for \texorpdfstring{$\Upsilon_{ij}$}{ECS}. Model d} \label{sec:new_model}

\begin{table*}[ht]
\caption{Stellar parameters.} 
\label{table:stellar_param}
\centering
\begin{tabular}{lcccc}
\hline\hline
\multicolumn{1}{c}{Param/Star} & Sun\ref{solar}  & HD 22049 & GJ 832    & GJ 581\\ \hline

Spectral Type   & G2 V  & dK2 V\ref{keenan}  & dM2 V\ref{houdebine}   & dM3 V\ref{trifonov} \\

$T_\textrm{eff}$ (K)    & 5772  & 5010 $\pm$ 64\ref{petit}  & 3590 $\pm$ 100\ref{houdebine}    & 3498 $\pm$ 56\ref{vonbraun11}\\

$R^\star/R_\odot$   & 1 & 0.74 $\pm$ 0.01\ref{baines} & 0.499 $\pm$ 0.017\ref{vonbraun11} & 0.299 $\pm$ 0.010\ref{vonbraun14}\\

$d$ (pc)    & 4.848e-6   & 3.2028 $\pm$ 0.0047\ref{gaia}   &   4.965 $\pm$ 0.001\ref{gaia}   & 6.30 $\pm$ 0.01\ref{gaia}\\

$\log\,\left[ \textrm{Fe}/\textrm{H} \right]^\star$  & 0    & -0.08 $\pm$ 0.04\ref{petit} & -0.06 $\pm$ 0.04\ref{lindgren}   & -0.33 $\pm$ 0.12\ref{bean}\\

$\log\,g$ (cgs) & 4.44  & 4.53 $\pm$ 0.08\ref{petit}  & 4.7\ref{schiavon}   & 4.92 $\pm$ 0.10\ref{bean}\\

Base Model Index & 1401\ref{fontenla} & 2227\ref{vieytes} & 3385\ref{tilipman} & 5812\ref{tilipman}  \\ \hline
\end{tabular}
\tablebib{
(\newtag{1}{solar}) \citet{solardata}; 
(\newtag{2}{keenan}) \citet{keenan:1989}; 
(\newtag{3}{houdebine}) \citet{houdebine:2016}; 
(\newtag{4}{trifonov}) \citet{trifonov:2018}; 
(\newtag{5}{petit}) \citet{petit:2021}; 
(\newtag{6}{vonbraun11}) \citet{vonbraun:2011};  
(\newtag{7}{baines}) \citet{baines:2012}; 
(\newtag{8}{vonbraun14}) \citet{vonbraun:2014};  
(\newtag{9}{gaia}) \citet{gaia:2018}; 
(\newtag{10}{lindgren}) \citet{lindgren:2017}; 
(\newtag{11}{bean}) \citet{bean:2006}; 
(\newtag{12}{schiavon}) \citet{schiavon:1997}; 
(\newtag{13}{fontenla})~\citet{fontenla:2015}; 
(\newtag{14}{vieytes}) \citet{vieytes:2021}; 
(\newtag{15}{tilipman}) \citet{tilipman:2021}
}
\end{table*}

The new \MgI model presented in this work (designated as ``d,'' in continuity with the models from Paper I) shares many of the characteristics of model ``c'' from the previously mentioned work (summarized in Table \ref{table:models}). The main differences between model d and c lie in the radiative and collisional data used to calculate NLTE populations.

\textbf{Calculation of radiative and collisional rates}. The radiative and collisional rates included in atomic models c and d were computed using the \textsc{autostructure} code \citep{badnell:2011}. The calculation considered a combination of 85 configurations in the LS coupled scheme, which included the $3s^2$ ground state, the single excited $3s\,nl$ states ($n \leq 20$), and the double excited states $3p\,nl$ ($n \leq 4$, $l \leq n-1$) and $3d^2$. In addition, all energy values for the terms were corrected with experimental measurements from the NIST database. The resulting atomic structure consists of 189 terms and 339 levels. We implemented the distorted-wave Breit-Pauli perturbative method (DW), which is included in \textsc{autostructure} to calculate electron impact excitation. From the collision strengths (CS), we derived the effective collision strengths ($\Upsilon_{ij}$) between levels $i$ and $j$, using the CS parametrization from \citet{burgess:1992}.

The atomic model designed for the radiative and collisional calculation (189 terms) differs from the models used in this work (85 terms). The energy structure of the 85-level\footnote{We follow the common nomenclature ``level'' to refer to the $^{2S+1}L$ term, and ``sublevel'' when referring to a fine-structure $^{2S+1}L_J$ level.} models, presented in Paper I and this work, was based on experimentally observed levels. In contrast, all theoretical levels are considered in our calculation. In addition, the 85-level models use superlevels, which are not formally included in the electronic structure theory. To resolve this, we modified the calculated theoretical structure (terms, levels, and radiative and collisional transitions) to match the 85-level models. We discarded nonobserved levels and combined terms in the proposed superlevels. We calculated the radiative and collisional transitions that use superlevels using the superlevel formalism proposed by \cite{anderson:1989}.

For the selection of spectral lines in NIST (described in Paper I), we used the condition of intense lines: $\log (\text{gf}) > -1$, which requires an atomic model with 54 detailed levels (extracted from the \textit{NIST}\footnote{\url{https://www.nist.gov/pml/atomic-spectra-database}} 5.7.1 database \citep{NIST:2020}) to represent them. This selection resulted in a model representing 285 spectral lines of \MgI (125 if only transitions between terms are considered). In the new model d, in addition to the spectral lines mentioned above, 5674 theoretical radiative transitions between bound states were included (3001 considering only transitions between terms). Figure~\ref{fig:histogram} shows a histogram with the number of term-term radiative transitions using a certain level. It can be seen that most of the spectral lines belonging to the FUV, NUV, and visible ranges are formed at relatively low levels. On the other hand, the lines belonging to the NIR and MIR use scattered levels throughout the atomic model. Figure~\ref{fig:histogram} also shows the importance of including, in addition, higher levels necessary for collisional and radiative population exchange between other energy levels and with the following ionization states. 

Regarding the collisional data, the main difference between models d and c is that in the new model, semiempirical methods of \cite{seaton:1962} and \cite{vanRegemorter:1962} (SEA\&VRM) are no longer employed to represent excitation by electron impact (previously included in model c for the transitions that involve levels 55 to 85 ). We replace it with the aforementioned DW calculations. We only consider DW values from or to terms equal to or greater than the term $3s\,6p\,^1P$, indexed as level 26.

Terms energy levels, effective collision strengths, and radiative data are available at the CDS.\footnote{Only available in electronic form at the CDS via anonymous ftp to \href{https://cdsarc.cds.unistra.fr}{cdsarc.cds.unistra.fr (130.79.128.5)} or via \url{https://cdsarc.cds.unistra.fr/cgi-bin/qcat?J/A+A/}
}

\begin{figure*}[ht]
\centering
\includegraphics[width=.5\textwidth]{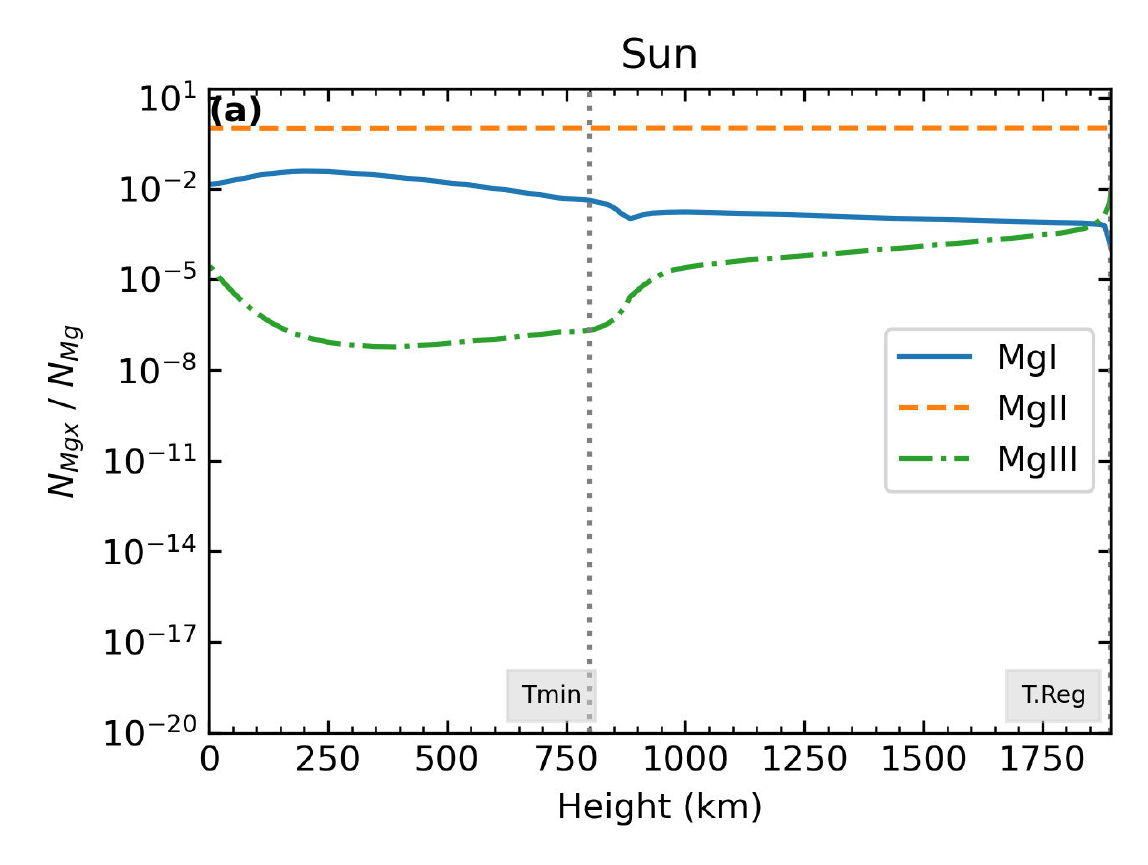}
\includegraphics[width=.5\textwidth]{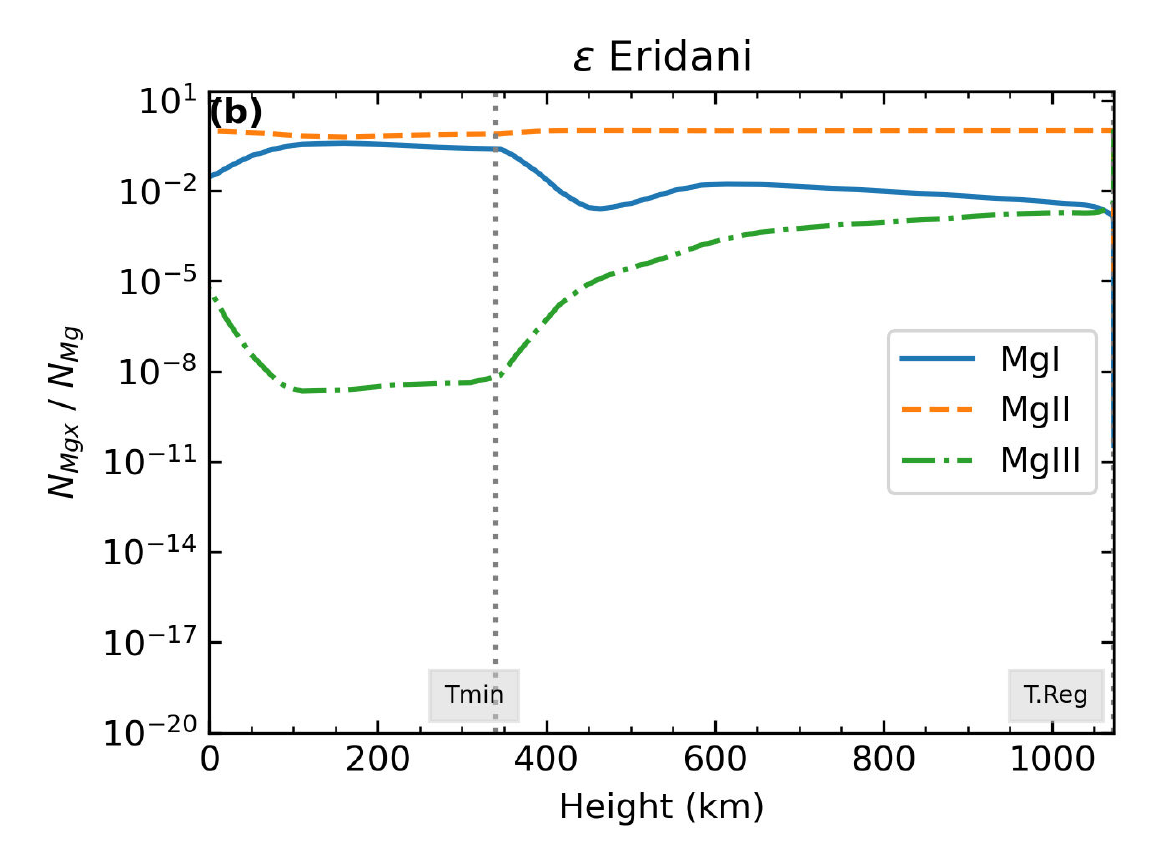}\vspace{0cm}
\includegraphics[width=.5\textwidth]{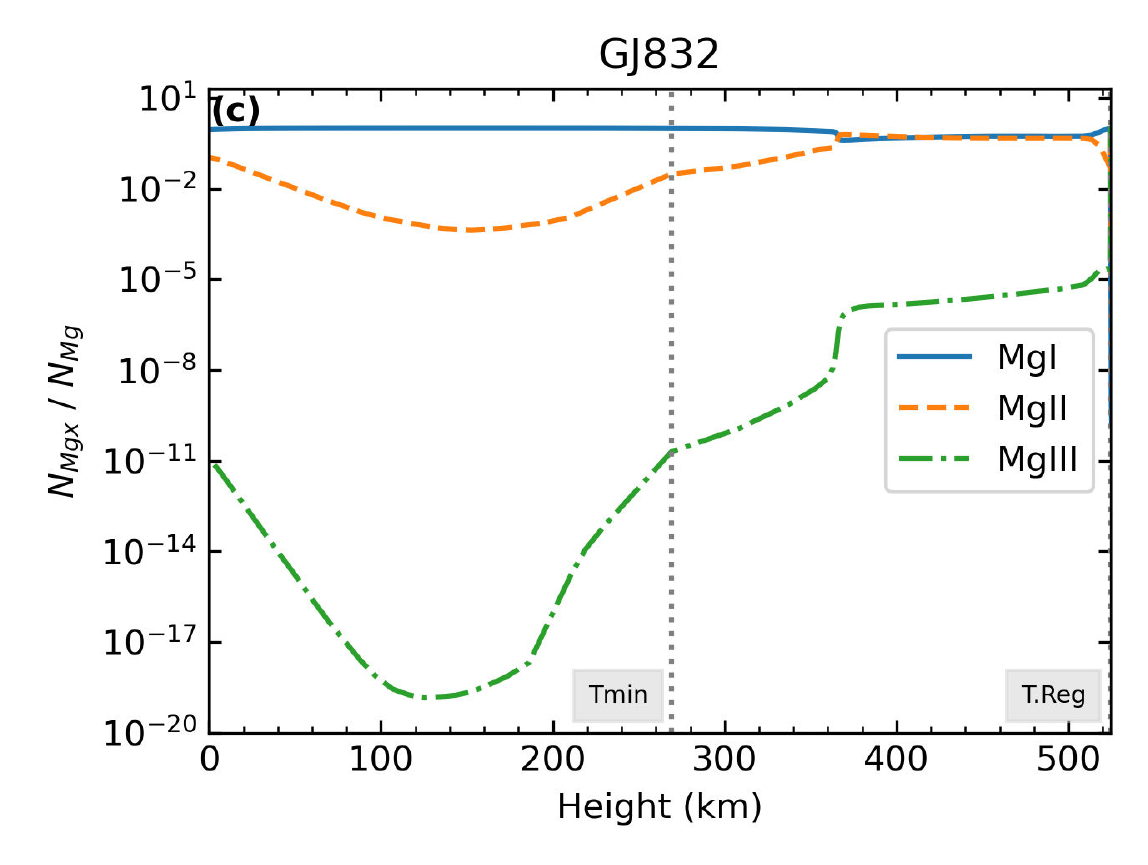}
\includegraphics[width=.5\textwidth]{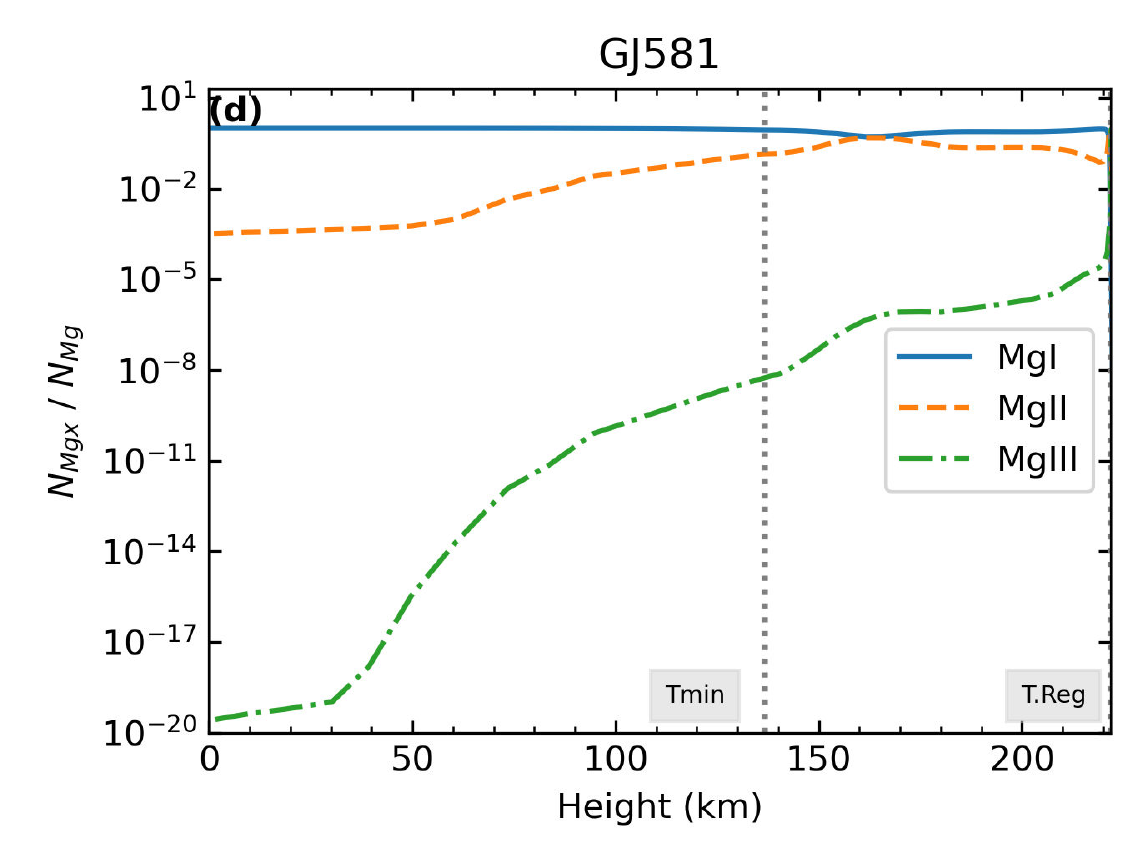}
\caption{Distribution of magnesium I (blue solid trace), II (orange dashed trace) and III (green dash-dotted trace) along the Sun's atmosphere (a), Epsilon Eridani (b), GJ 832 (c) and GJ 581 (d) using the base model. The heights at the temperature minimum (``Tmin'') and the beginning of the transition region (``T.Reg'') for each star are noted in light grey boxes and dotted lines.}
\label{fig:mg_distribution}
\end{figure*}

\section{Atmospheric models: The Sun and three cooler stars} \label{sec:atmos_models}

\begin{figure*}[ht]
\centering
\includegraphics[width=.5\textwidth]{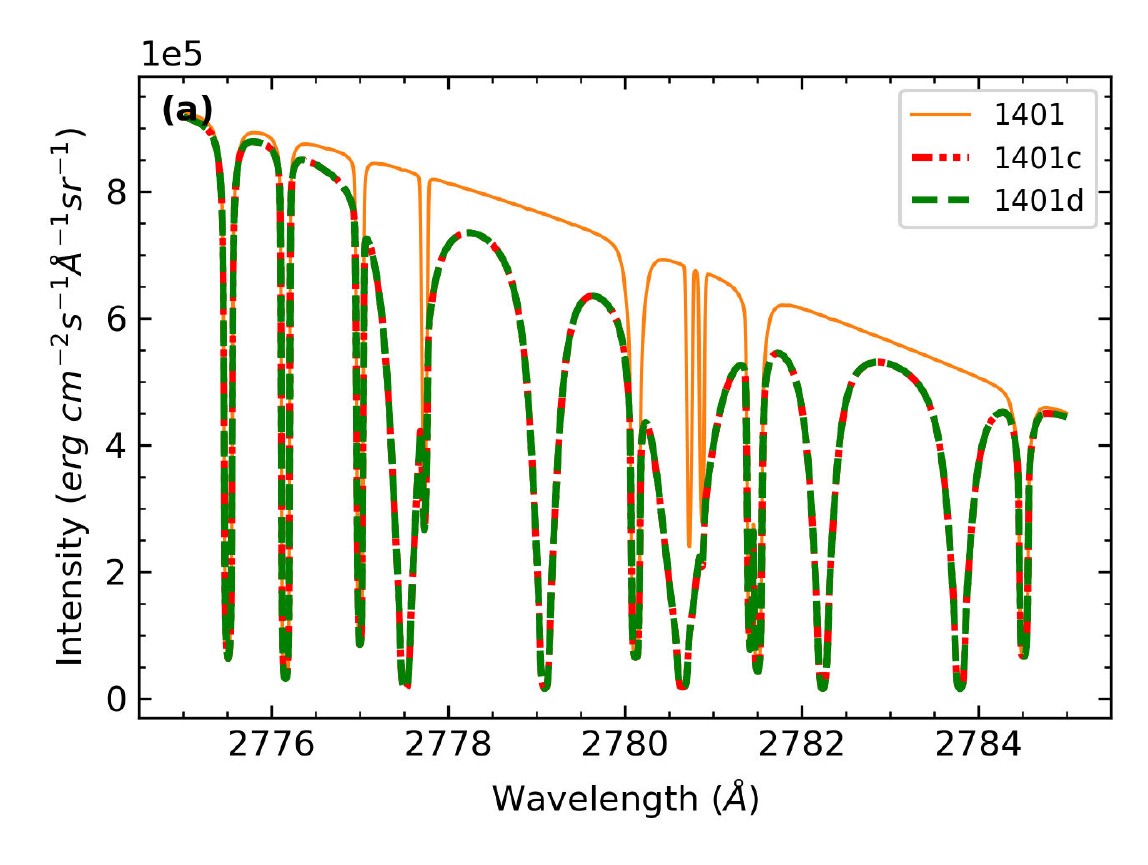}
\includegraphics[width=.5\textwidth]{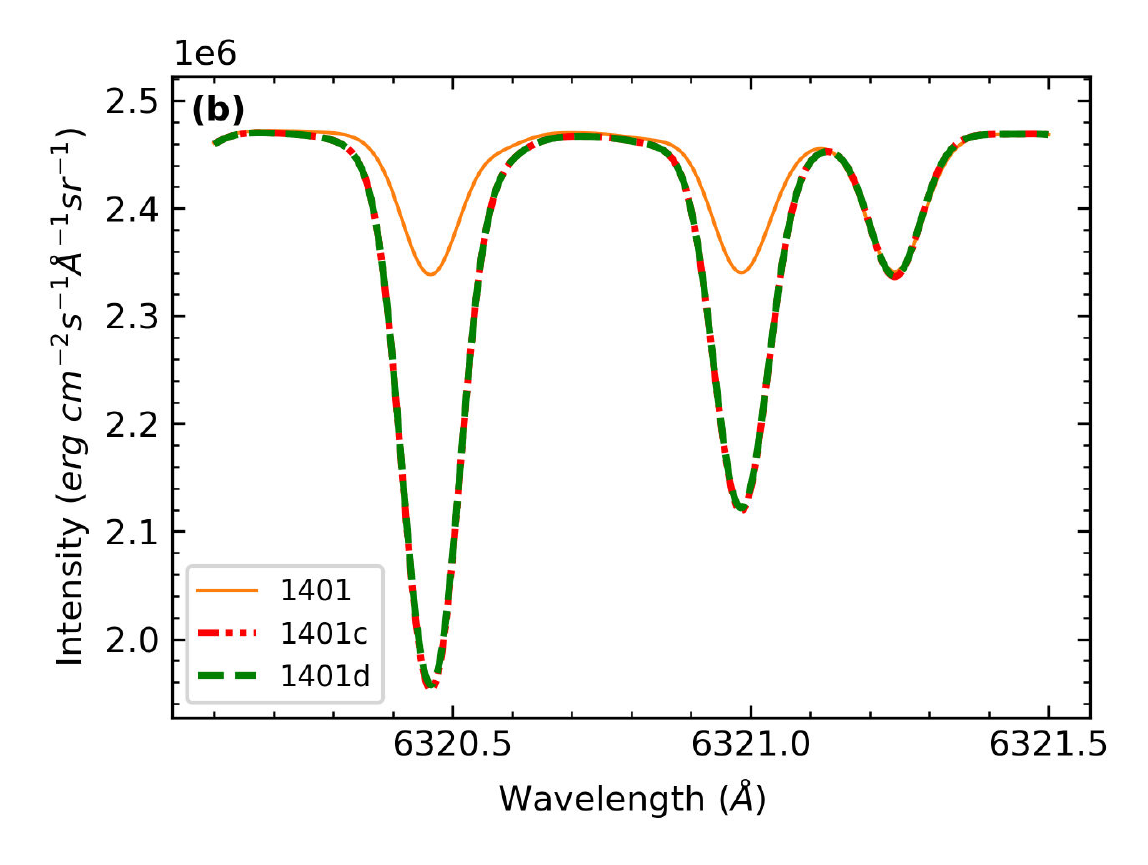}\vspace{0cm}
\includegraphics[width=.5\textwidth]{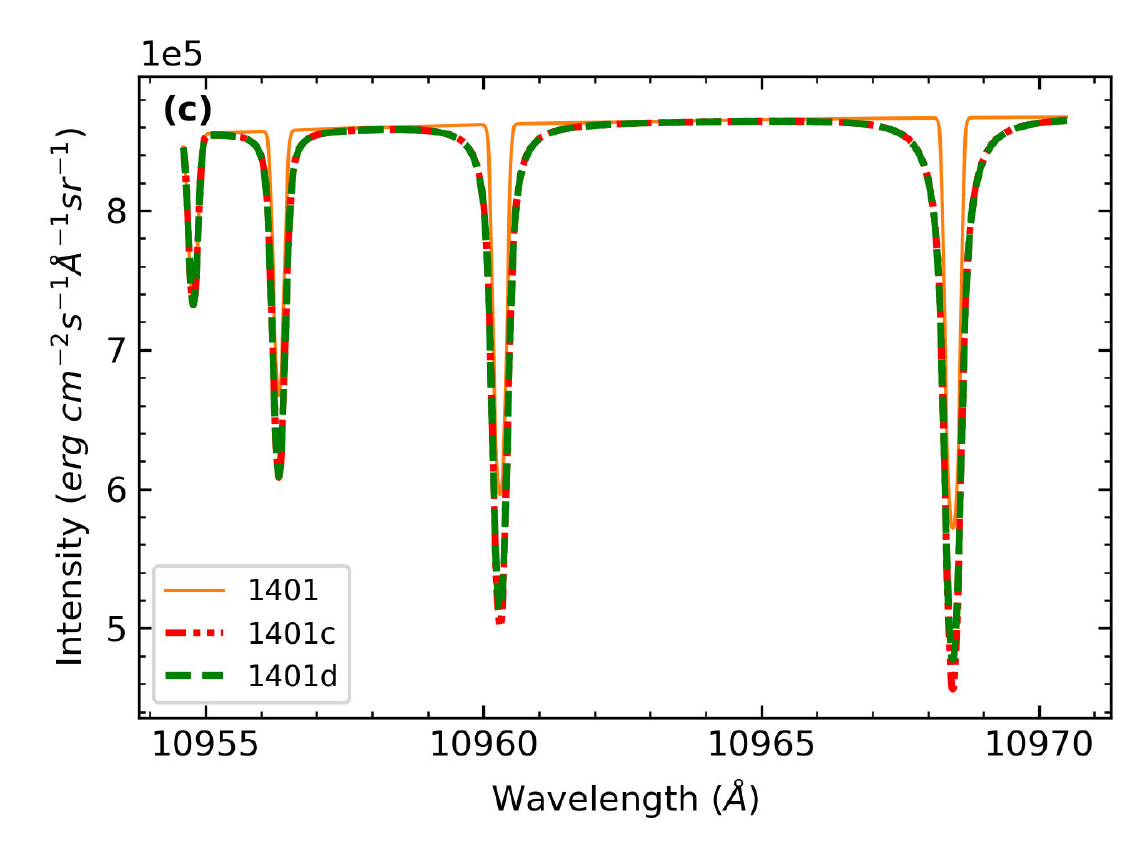}
\includegraphics[width=.5\textwidth]{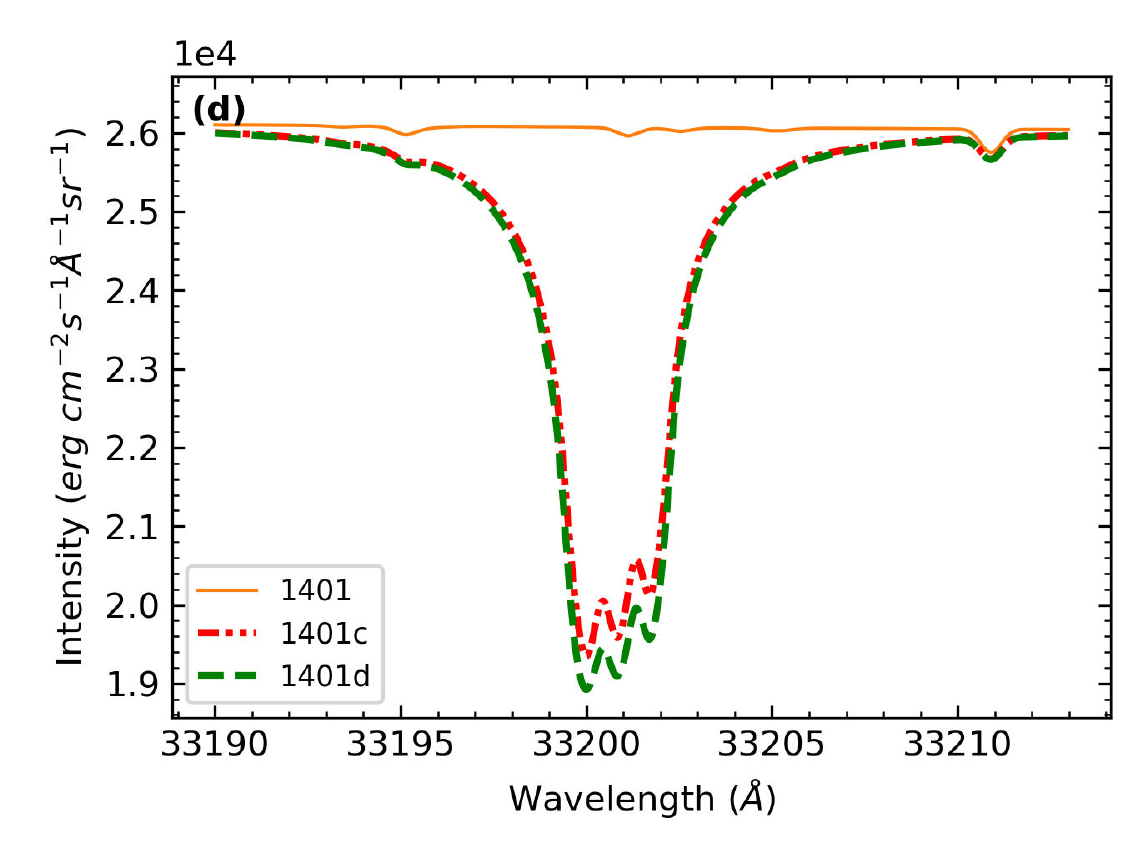}
\caption{Some of the lines presented in Paper I (model c) representative of each spectral range. Comparison between the models: base (continuous orange line), c (dash-dot-dot red line), and the new model d (dashed green line).}
\label{fig:sun}
\end{figure*}

Unlike atmospheric model grids, which are constrained by theoretical predictions, semiempirical models built with the SSRPM have as input a large number of observations in different spectral ranges. The degrees of freedom of the model are given by the need for flux-calibrated observations covering the largest possible range of wavelengths. A good semiempirical model should be able to fit all possible observations in order to be reliable. In this way, it is possible to predict regions of the spectrum that cannot be obtained by direct observations. The model is based on a set of element abundances that conform the stellar atmosphere, and a grid of heights on which calculations are performed. Each height is characterized by its temperature ($T$), microturbulent velocity ($v_t$), and numerical densities of protons ($n_p$), electrons ($n_e$), total hydrogen atoms, and neutral hydrogen atoms ($n_a$). In other words, our input parameters are given only by the metallicity, the surface gravity (usually obtained from literature), and the aforementioned densities.

Assuming an atmospheric model to be correct, a reliable atomic model should correctly reproduce the observed spectral lines. In addition, it should also be able to correctly reproduce the observed line profiles when switching to another atmospheric model with different plasma parameters. To test the reliability of the new atomic model of \MgI (d), we used atmospheric models of stars of different spectral types, which were previously calculated with the SSRPM system. These models are briefly described below and illustrated, from the photosphere to the transition region, in the left panel of Fig.~\ref{fig:T_ne_vs_press} as a function of pressure. The stellar parameters of the present stars are summarized in Table~\ref{table:stellar_param}.

\textbf{Sun}. This atmospheric model is the so-called 1401 in \citet{fontenla:2015}. It represents the most abundant structure on the solar disk in a quiescent state and was used in Paper I. For the current work, we will analyze the solar spectrum in the same way as the stars, that is to say integrated on the disk.

\textbf{GJ 832 and GJ 581}. GJ 832 was introduced as 3346 in \citet{fontenla:2016}. It was later improved by \citet{tilipman:2021} (labeled as 3385) to correct an overestimation of the flux in the visible range due to incorrect calibration of the observations used to generate the model. The atmospheric model of GJ 581 is also published in the mentioned paper. In this type of cool star, the extremely high density of molecular lines formed in the photosphere makes it difficult to distinguish the continuum level. The intensity of the pseudo-continuum is given at a temperature where $\tau=2/3$. This value corresponds to the stellar emission source because, since stars cannot be observed with spatial resolution, we observe the emerging flux for an average angle of $\mu=\cos \theta = 2/3$. The authors note that, although the shape and intensity of the flux are correctly reproduced by the models, there is an overestimation of the flux in the visible range, probably due to the absence of a source of molecular opacity between 4000 and 4500 {\AA}. 

\textbf{Epsilon Eridani (HD 22049)}. \citet{vieytes:2021} published a developing model for this star. The atmospheric model includes the photosphere, chromosphere, transition region, and corona. In the visible range, the model correctly reproduces the observed continuum and important lines that characterize the chromospheric magnetic activity. In the FUV and EUV regions, the model reproduces lines and the continuum that cannot be easily observed due to interstellar absorption. However, the authors mention that the thermal structure of the corona still needs to be modified to include observations in the X-ray range. In this work, we use the model (denoted as 2227) from the photosphere to the beginning of the transition region, together with the previously mentioned models.

As observed in the left panel of Fig.~\ref{fig:T_ne_vs_press}, the solar model 1401 presents a higher temperature than the other stars in the photosphere (from right to left). This situation is exchanged with model 2227 for the star Epsilon Eridani around $5 \times 10^2\, \text{erg}/\text{cm}^2$. The temperature minimum in the last model occurs at higher pressure, and therefore the first chromospheric rise occurs earlier. The temperature of its chromospheric plateau and the rise to the transition region also occur at higher pressure and temperature, indicating that this star is more active than the Sun. In the case of the 3385 and 5812 models of the stars GJ 832 and GJ 581, respectively, it can be seen that they are much cooler and are similar to each other up to their temperature minimum. From this point, as the pressure decreases, while 3385 has a steeper first chromospheric rise, 5812 has a more gradual temperature rise and, consequently, a relatively smaller chromospheric plateau until it reaches the temperature rise that marks the beginning of the transition region. These last two stars are optically not very active compared to Epsilon Eridani. 

In the right panel of Fig.~\ref{fig:T_ne_vs_press}, the electron density ($ne$) is plotted as a function of the pressure for each star. This figure facilitates the analysis of the importance of the e+\MgI collisional process in the region of interest. Moving through the figure from right to left, it can be seen in all cases that the electron density decreases as we move away from the photosphere up to regions close to the temperature minimum. This behavior is expected since the average density of most of the components in the atmosphere decreases together. This decrease is reverted when the chromospheric rise is reached, where a noticeable increase in $ne$ as a product of the ionization of metals starts. 

Concerning the magnesium in each atmosphere, Fig.~\ref{fig:mg_distribution} plots the atmospheric distribution of the \MgI, \MgII, and \MgIII density to the total Mg density. As all figures are on the same scale, the differences can be clearly identified. In Epsilon Eridani, the distribution is similar to the solar distribution, with \MgII predominating throughout the atmosphere. Between 50 km and the temperature minimum ($T_{\text{min}}$), the population of \MgI and \MgII is almost of the same order. For altitudes above $T_{\text{min}}$, the \MgI population decreases due to the increase in the ionization rate in this region, an effect that is also visible through the increase in \MgIII in a correlated manner. The predominance of \MgII is the main reason we use near-continuous superlevels, which allows population exchange between \MgI and \MgII. In the case of dM stars, the distribution is very different, with neutral magnesium predominating as the main component up to a few kilometers above $T_{\text{min}}$. The lower temperatures do not favor the ionization process, and the population of \MgII and \MgIII is drastically reduced by recombination. 

Quantitative results of this distribution are presented in Section \ref{subsec:d_populations}

\section{Observations} \label{sec:observations}

\begin{table*}[t]
\caption{Observational data used in this work}
\label{table:observations}
\begin{tabular}{lccclc}
\hline\hline
\multicolumn{1}{c}{Instrument} & Res. Power           & Obs. Label           & Star            & \multicolumn{1}{c}{Obs. Date}               & Fig.       \\
\hline
Stratospheric Baloon\ref{ha}           & 50000                & H\&A                 & Sun             & 1978-04-20+                      & 10, 13     \\
STIS. E230H gr. at HST\ref{muscles}     & 114000               & STIS-E230H           & Epsilon Eridani & 2015-02-01 0:42:55                          & 10         \\
STIS. G230L gr. at HST\ref{muscles}     & 500-960              & STIS-G230L           & GJ 832          & 2014-10-10 4:01:50                          & 10         \\
                               & 500-960              & \multicolumn{1}{l}{} & GJ 581          & 2015-08-10 22:06:10                         & 10         \\
COS. G230L gr. at HST\ref{muscles}      & 1550-2900            & COS-G230L            & GJ 832          & 2014-10-10 11:45:42 & 10         \\
                               & 1550-2900            &                      & GJ 581          & 2015-08-11 7:17:30                          & 10         \\
STIS. E230M gr. at HST\ref{muscles}     & 30000                & STIS-E230M           & Epsilon Eridani & 2015-02-01 0:26:16                          & 13         \\FTS Solar Atlas at Kitt Peak\ref{kpfts}                & 350000               & KP-FTS               & Sun             & 1962+                      & 14, 15, 16 \\
ESPRESSO VLT-ESO\ref{espresso}               & 70000-190000         & ESPRESSO             & Epsilon Eridani & 2018-10-28 3:08:50                          & 14         \\
HARPS-ESO\ref{harps}                      & 115000               & HARPS                & GJ 832          & 2020-01-01 1:00:40                          & 14         \\
                               & 115000               & \multicolumn{1}{l}{} & GJ 581          & 2012-05-14 4:09:45                          & 14         \\
UVES-ESO\ref{uves}                       & 107200               & UVES                 & Epsilon Eridani & 2012-12-17                                  & 15         \\
                               & 42310                & \multicolumn{1}{l}{} & GJ 581          & 2006-06-06 23:27:21                         & 15         \\
FEROS-ESO\ref{feros}                      & 48000                & FEROS                & GJ 832          & 2012-08-06 3:00:29                          & 15         \\
                               & \multicolumn{1}{l}{} & \multicolumn{1}{l}{} & GJ 581          & 2015-02-19 8:17:35                          & 15\\        
FTS Solar Atlas at SCISAT-1\ref{acefts}                      & 50000                & ACE-FTS                & Sun          & 2004+                          & 17, 18         \\      
\hline\hline
\end{tabular}
\tablefoot{The values presented as reference in the Power Res. column were extracted from the .fits file when available, and from the instrument page otherwise.}
\tablebib{
(\newtag{1}{ha}) \citet{hall:1991}; 
(\newtag{2}{muscles}) \citet{loyd:2016}; 
(\newtag{3}{kpfts}) \citet{neckel:1999}; 
(\newtag{4}{espresso}) \citet{ESPRESSO:2020}; 
(\newtag{5}{harps}) \citet{HARPS:2003}; 
(\newtag{6}{uves}) \citet{UVES:2000}; 
(\newtag{7}{feros}) \citet{FEROS:1999}; 
(\newtag{8}{acefts}) \citet{acefts:2010}; 
}
\end{table*}

In addition to comparing computed spectra between models, we included astronomical observations whenever possible. For the Sun, we used the observations described in Paper I: for the NUV range, the observations of \citet[H\&A hereafter]{hall:1991}; for the visible and NIR, the observations from the Fourier-Transform-Spectra Solar Atlas (KP-FTS hereafter) obtained at Kitt Peak Observatory \citep{neckel:1999}; and for the MIR, the transmittance data from the ACE-FTS Solar Atlas \citep{acefts:2010}, recorded by the spectrometer aboard the spacecraft SCISAT-1. The observational search for the stars' spectra was extensive, with special attention being paid to the UV and IR ranges (regions where we had the most prominent changes between models). However, observations with sufficient resolution and high signal-to-noise ratio were difficult to find and even nonexistent in some cases. We describe below the observation data used in this work for the stars Epsilon Eridani, GJ 832, and GJ 581.

\textbf{FUV \& NUV}. In this range, we included the spectra obtained by the MUSCLES Treasury Survey program\footnote{\url{https://archive.stsci.edu/prepds/muscles/}} \citep{loyd:2016} through the Hubble Space Telescope (HST), mainly the instruments of the Space Telescope Imaging Spectrograph (STIS). We used the medium-resolution E230M grating (STIS-E230M hereafter) and the high-resolution E230H grating (STIS-E230H hereafter) for Epsilon Eridani. In the FUV, we did not find observations of GJ 832 and GJ 581 that resolved such thin lines ($\sim$0.5 {\AA}), and we could only add observations of Epsilon Eridani (STIS-E230M) in one of the examples. In the NUV, for the dM stars GJ 832 and GJ 581, we employed the low-resolution G230L grating (STIS-G230L hereafter). These data were used to construct the respective stellar models in \cite{tilipman:2021} and \cite{vieytes:2021}. In addition, for the 2853.0 {\AA} line of the dM stars, we also used HST observations obtained by the Cosmic Origins Spectrograph Instrument (COS-G230L hereafter) using the G230L grating. These additional observations were also obtained by the MUSCLES program on the same dates as the previous ones, so they are considered relevant for comparison.

\textbf{VISIBLE}. The atmospheric models used in this work were constructed with the visible range continuum level provided by observations at Complejo Astronómico El Leoncito (CASLEO) Observatory. However, they are not included in the figures since they lack sufficient resolution to reproduce the selected lines. Therefore, for Epsilon Eridani, we selected the most recent and highest resolution observation (0102.D-0185(A)) taken with the ESPRESSO spectrograph, part of the Very Large Telescope (VLT) of the European Southern Observatory (ESO). For GJ 832 and GJ 581, we used the latest observations (0104.C-0863(A) and 183.C-0437(A), respectively) made with the high-resolution HARPS spectrograph, installed on ESO's 3.6 m telescope.

\textbf{NIR \& MIR}. In this range, for both Epsilon Eridani and GJ 581, we use the most recent and highest signal-to-noise ratio observations (090.D-0039(A) and 077.D-0066(A), respectively) from the high-resolution UVES spectrograph, located at the Nasmyth B focus UT2 of the VLT, at ESO. For GJ 832 and GJ 581, we used the latest data (089.C-0440(A) and 094.A-9029(I), respectively) from the FEROS spectrograph, installed on the 2.2 m MPG/ESO telescope at ESO's La Silla Observatory. Except for the Sun, we could not find observational data for the stars with sufficient spectral resolution at MIR.

Air-to-vacuum wavelength conversions and Doppler shifts were performed. For the latter, the radial velocity of the object was used according to The SIMBAD astronomical database \citep{SIMBAD:2000}. Like the SRPM, the SSRPM produces spectra calculated with a spectral resolution of $R=10^6$. For a proper comparison with observations, the calculated spectra were convolved with a Gaussian function to reduce their resolution and match those of the corresponding instrument. In addition, for the spectra observed at line 2853.0 {\AA}, the flux received on Earth was increased by 30\% to account for the absorption from the interstellar medium \citep{france:2013}. Then, it was converted to flux at the surface of the star, using the relation: $F^\star= \pi \, (R^\star/d)^2$, with the distances and radii presented in Table~\ref{table:stellar_param}. Regarding the flux in the other observations, it should be noted that the continuum of the selected observations was normalized to that of the CASLEO observations (and, therefore, to that of the calculated spectra). For this reason, the spectra are not represented on an absolute scale. Instead, they are matched to their continuums in a wider range and normalized to the local maximum. Finally, due to the lack of observations, the lines calculated in the MIR range of the stars were convolved to the ACE-FTS resolution for comparison purposes only.

Table \ref{table:observations} contains details of the observations used and other information of interest.

\section{Results and discussion} \label{sec:results}

We expand the study carried out in Paper I for the Sun. Particularly, we analyze the effect produced by the change in the atomic parameters of \MgI in the atmosphere of stars of different spectral types. It should be noted that when performing the NLTE calculation with the d model, the atomic populations vary as a result of all the changes introduced to the base model (described in Section \ref{sec:new_model}). Obtaining a synthetic spectrum that agrees with the observations is important. However, even if there are no appreciable variations in the spectral lines, we can analyze in detail the changes in the energy level populations by inspecting the atmospheric models. Thus, it is possible to perform a comprehensive analysis that can be generalized to other atmospheric conditions, as in the case of stars of different spectral types, metallicities, levels of activity, etc. To this end, we compare the new model d against the base model for each star.

In continuity with Paper I, we first present the results produced by the new model d on the solar atmospheric model. As explained in the previous Section, the semiempirical formulas of SEA\&VRM are completely discontinued in model d for the calculation of $\Upsilon_{ij}$. We replace these values with DW calculations for energy levels from 26 ($3s5g\,^1G$, 57\,262.76 $cm^{-1}$) up to, and including, level 85 ($3p^2\,^1S$, 68\,275 $cm^{-1}$). The main characteristics of the model are summarized in Table \ref{table:models}.

\subsection{Model d versus c, for the solar spectrum} \label{subsec:d_vs_c}

In this section, we present the differences found for \MgI when comparing the results obtained with the present model d and c from Paper I. In this way, we give continuity to the study previously carried out with an improved calculated model, conformed by more reliable collisional data.

\begin{figure}[t]
\centering
\resizebox{\hsize}{!}{\includegraphics{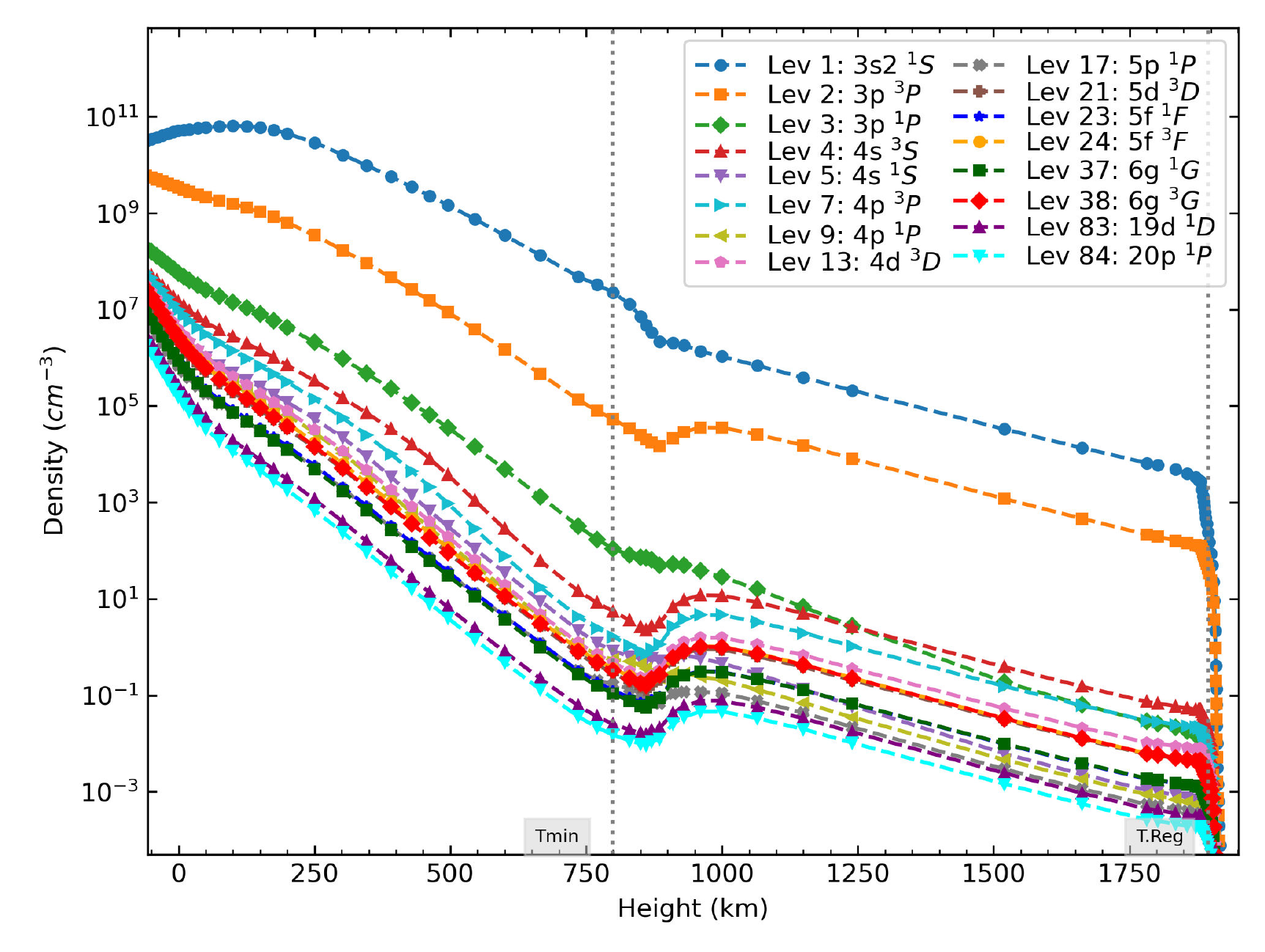}}
\caption{Distribution of \MgI at energy levels of interest (model d) as a function of height from the solar photosphere to the transition region.}
\label{fig:density_h_d_sun}
\end{figure}

Figure~\ref{fig:density_h_d_sun} illustrates the population density of \MgI for model d as a function of height (up to before the transition region) for certain levels of interest. Particularly, we show the lower and higher levels (and those in between) of the transitions that correspond to the lines we analyze in the following sections. This figure shows the different orders of magnitude in the energy levels populations and the density variation as we move to different regions of the atmosphere (solar, in this case). It can be seen that the lower-lying levels present the highest population density (e.g., in model d: 91.54 \% is in level 1, 6.86 \% in level 2, and 0.18 \% in level 3), and at low altitudes of the atmosphere ($\lesssim$ 100 km). One should bear in mind that the effect of modifying certain collisional parameters on a given spectral line will depend on the population variation in the levels involved, as well as on the characteristics of the atmosphere in the region where the line is formed. Thus, for the same population change in the corresponding levels, the formation of certain lines will be more affected than others. Moreover, some lines can even be more strongly affected than the same lines in other stars. Examples of these cases will be shown in Section \ref{subsec:lines}.

Figure~\ref{fig:sun} shows four lines as a representation of the different regions of the solar spectrum, calculated with the models: base, c (from Paper I), and the present model d. We selected these lines as they were previously examined in Paper I. It can be seen that models c and d produce identical lines in the NUV (Fig. \ref{fig:sun}a) and VIS (Fig. \ref{fig:sun}b) regions. In contrast, in the NIR (Fig. \ref{fig:sun}c) and MIR (Fig. \ref{fig:sun}d) regions, small absorption differences can be distinguished. These differences are present, with varying amplitudes, throughout the NIR and MIR. We recall that the thermal structure of the solar atmospheric model, which is employed for the 1401c and 1401d calculations, was built assuming the atomic data of the base model (1401). This model features outdated \MgI atomic data (as detailed in Paper I) and only includes lines up to 17\,000 {\AA}. Nevertheless, the atmospheric models presented in this paper correctly represent the continuum and the most important lines of the stars in question. Hence, a follow-up work should include a new atmospheric thermal structure built with the present atomic model to correctly fit the observations (up to 71\,500 {\AA} in our case).

\subsection{The effect of the model d in the atomic populations} \label{subsec:d_populations}

It is important to keep in mind that the stellar atmosphere is a coupled system of macroscopic scale effects (e.g., through the radiation field originating at different heights) with atomic scale effects (e.g., through photon absorption and scattering processes that affect the material properties). In NLTE conditions, these effects are described by the system of equations consisting of the radiation transport, statistical equilibrium, and hydrostatic equilibrium equations. Hence, it is possible to understand that causally isolating the effects of modifying an atomic parameter in a spectral line is of great complexity in most cases. However, a detailed analysis can be performed by studying the populations of the energy levels that form the line and the atmospheric conditions in the formation region (as explained in Paper I). To obtain a more general notion of the spectral lines formation due to an atomic model in a certain star, we can compare how the populations are modified for a known model. Without looking at spectral lines, it is possible to obtain a general and complete idea of the differences between models by studying the distributions of the element of interest in its different ionization states, energy levels, or atmospheric heights, as required. In this Section, we present the changes produced in the atomic populations of magnesium with respect to the base model due to the improvements incorporated in the new model d. The indices used for each atmospheric model are detailed at the bottom of Table \ref{table:stellar_param}. 

\begin{table*}[ht]
\caption{Magnesium population distributions, and comparison over 26-levels base model.} 
\label{table:mg_distrib}
\centering
\begin{tabular}{lcccc c ccc}
\hline\hline
\multirow{2}{*}{Index} & \multicolumn{4}{c}{Distribution\tablefootmark{(a)} (\%)} && \multicolumn{3}{c}{Difference\tablefootmark{(b)} (\%)} \\
\cline{2-5} \cline{7-9}
                       & \MgI   & \MgII  & \MgIII   & Mg mol   && \MgI   & \MgII  & \MgIII         \\ \hline
1401                   & 1.47   & 98.46  & 7.8e-02  & 7.2e-05  &&                 &                 &                \\
1401d                  & 1.50   & 98.42  & 7.8e-02  & 7.3e-05  && 2.6e+00         & -3.8e-02        & -4.6e-01       \\ \hline
2227                   & 4.65   & 95.32  & 3.3e-02  & 5.3e-04 &&                 &                 &                \\
2227d                  & 4.74   & 95.22  & 3.3e-02  & 5.4e-04  && 2.0e+00         & -9.7e-02        & -7.0e-01       \\ \hline
3385                   & 72.20  & 27.62  & 2.9e-06  & 1.8e-01  &&                 &                 &                \\
3385d                  & 72.05  & 27.76  & 2.9e-06  & 1.8e-01  && -2.0e-01        & 5.2e-01         & 2.3e-01        \\ \hline
5812                   & 91.37  & 8.50   & 6.8e-06  & 1.3e-01  &&                 &                 &                \\
5812d                  & 91.20  & 8.67   & 6.9e-06  & 1.3e-01  && -1.8e-01        & 2.0e+00         & 5.1e-01 \\
\hline
\end{tabular}
\tablefoot{
\tablefoottext{a}{Relative to a Mg abundance of $Mg/H = 2.88e-05$.}
\tablefoottext{b}{Computed as: $100\cdot(M_d^{star} - M_{base}^{star}) / M_{base}^{star}$}
}
\end{table*}

\textbf{Magnesium distribution through its ionization states}. For a general understanding of the behavior of the element in each star, it is essential to know how the element is distributed in its main ionization states. Table \ref{table:mg_distrib} shows the distribution of magnesium in its most abundant states (\MgI, \MgII, \MgIII) and the fraction that forms molecular compounds for the base and d models. It is possible to verify quantitatively, and in agreement with Fig.~\ref{fig:mg_distribution}, the predominance of \MgII (greater than 95 \%) in the atmosphere of the Sun (prefix 1401) and Epsilon Eridani (prefix 2227); and the majority of \MgI (greater than 72 \%) in the cooler atmospheres of GJ 832 (prefix 3385) and GJ 581 (prefix 5812). It is also noteworthy  the amount of magnesium that forms molecules in each case.  In the Sun and Epsilon Eridani, there are as much as four orders of magnitude less molecular Mg than \MgI, while in the models for GJ 832 and GJ 581, the amount of molecular Mg is less than two orders of magnitude smaller than of \MgI. If we compare it with the hotter stars, the presence of molecular Mg is at least three orders of magnitude larger. These values become even more relevant when comparing the formation of the same line in different 
stars. Moreover, in many cases, lines formed in a dM star may end up blended by the molecular bands.
Examples of these cases will be shown in Section \ref{subsec:lines}.

We also analyze the differences in the total population of each star due to the models base and d. By inspecting the ``Difference'' column
of Table \ref{table:mg_distrib}, the behavior of the hottest and coolest stars can be grouped. For the Sun and Epsilon Eridani, model d shows an increase in \MgI greater than 2\%, correlated with a decrease in \MgII and \MgIII, with respect to each of their total densities. This result suggests a migration of \MgII toward \MgI, probably through recombination processes. Such processes are favored in model d because it features a more extensive \MgI electronic structure (with energy levels closer to the continuum) than the model base.
This argument is reinforced by comparing the same models in cooler stars, where for both GJ 832 and GJ 581, the migration occurs from \MgI to \MgII and \MgIII. In this case, ionizing collisions can be accounted as responsible for this phenomenon; in model d, these collisions may be favored due to the small energy gaps.

\begin{figure*}[ht]
\centering
\includegraphics[width=0.8\textwidth,keepaspectratio]{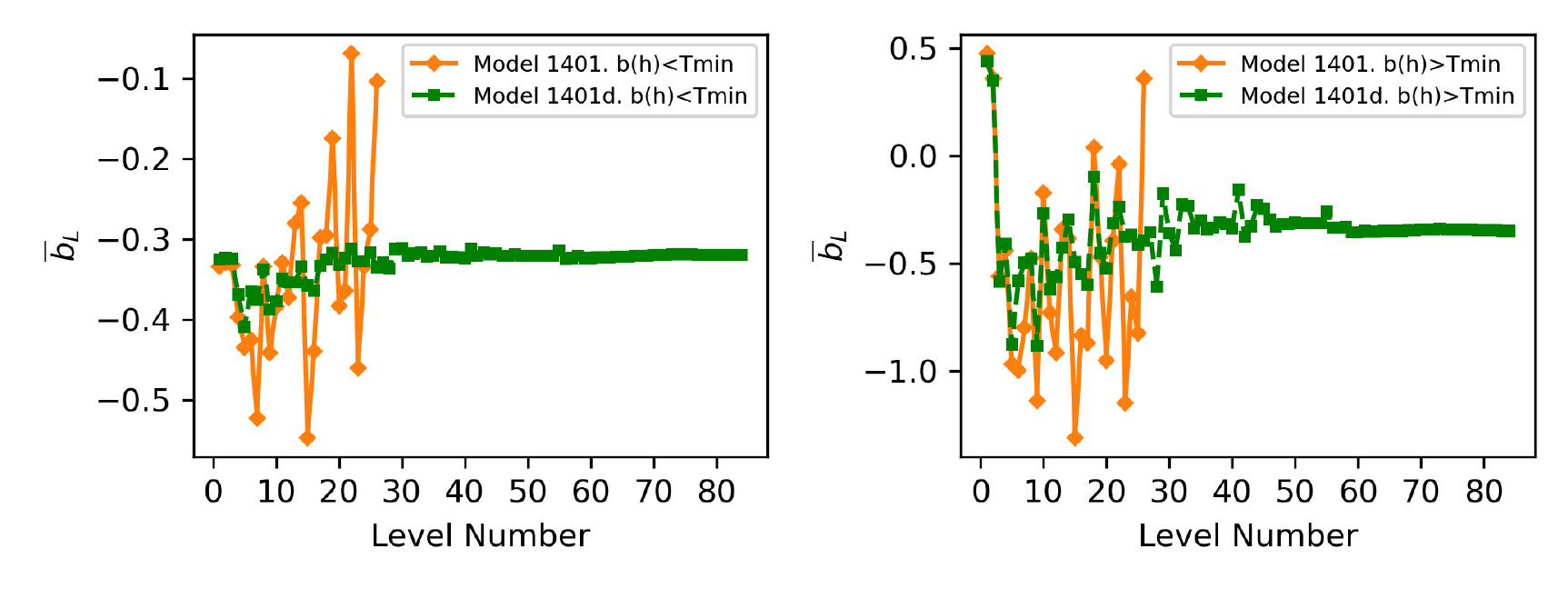}\vspace{0cm}
\includegraphics[width=0.8\textwidth,keepaspectratio]{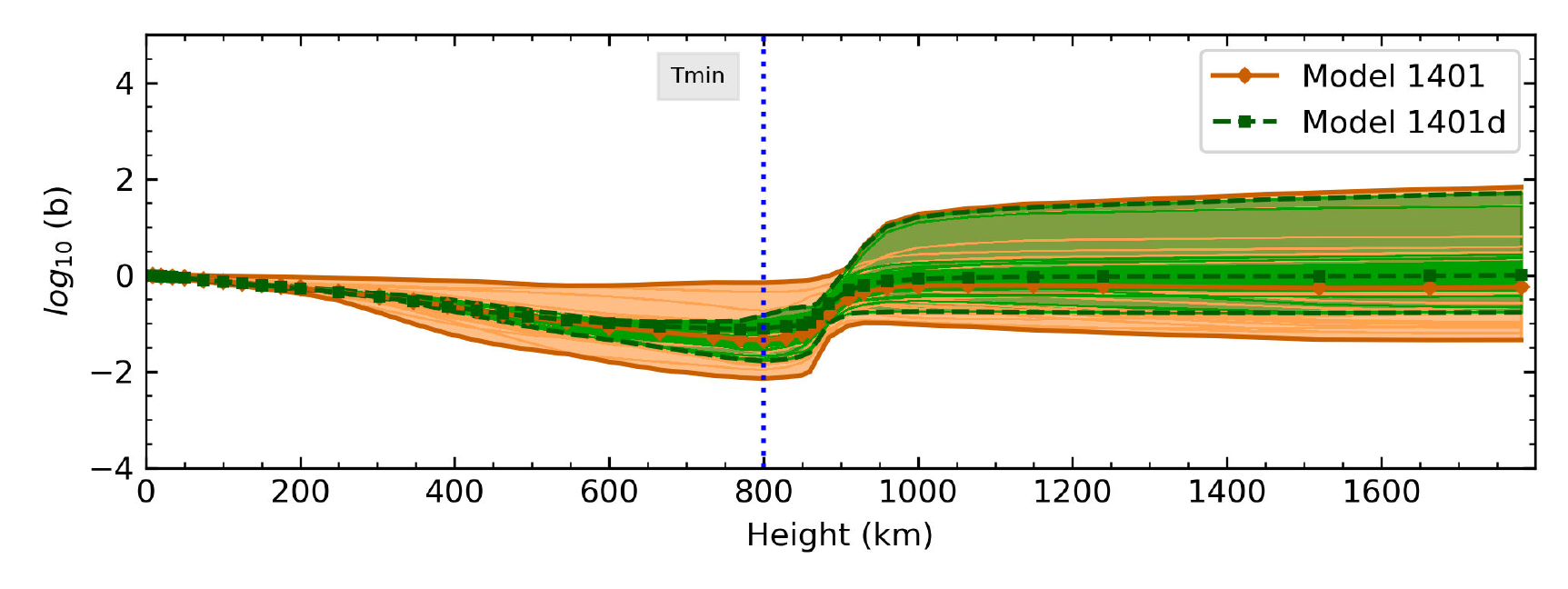}
\caption{Comparison of the distribution of \MgI in its energy levels (upper panel) and atmospheric heights (lower panel) between the 26-level base model (in orange) and the new 85-level model d (in green) for the Sun. The top panel shows the height-averaged LTE departure coefficient $\overline{b_L}$ up to the temperature minimum (top left panel) and from the temperature minimum to before the transition region (top right panel), as a function of the energy level number. In the lower panel, all the separation coefficients of each model are plotted as a function of height.}
\label{fig:b_Sun}
\end{figure*}

\begin{figure*}
\centering
\includegraphics[width=0.8\textwidth,keepaspectratio]{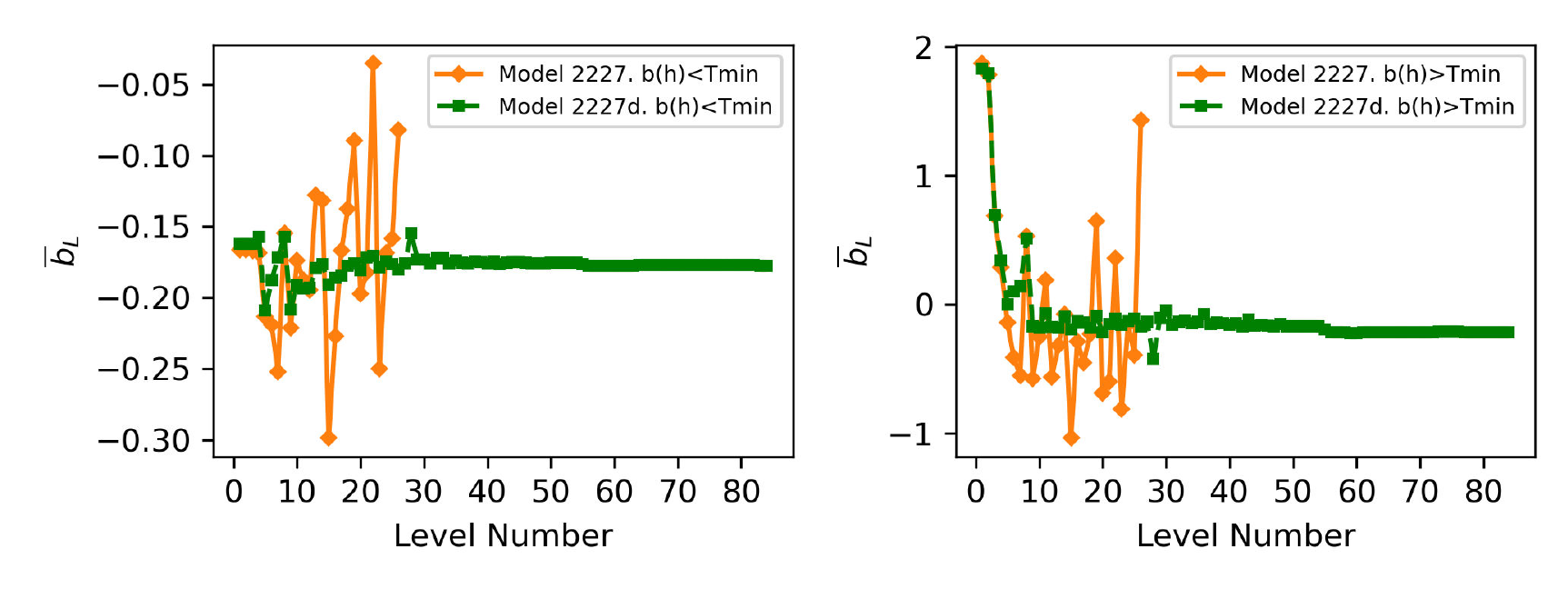}\vspace{0cm}
\includegraphics[width=0.8\textwidth,keepaspectratio]{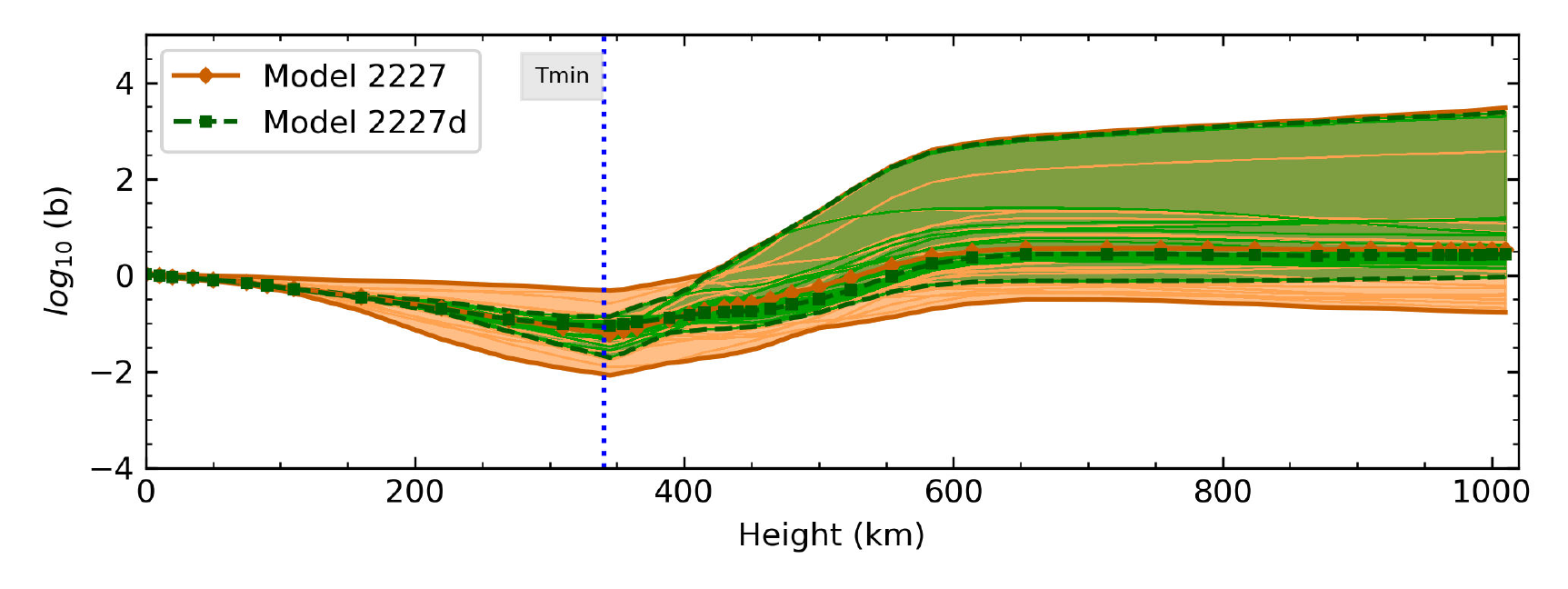}
\caption{Comparison of the distribution of \MgI in its energy levels (upper panel) and in the heights of the atmosphere (lower panel) between the 26-level base model (in orange) and the new 85-level model d (in green) for Epsilon Eridani. Details of subplots equal to Fig.~\ref{fig:b_Sun}.}
\label{fig:b_EERI}
\end{figure*}

\begin{figure*}
\centering
\includegraphics[width=0.8\textwidth,keepaspectratio]{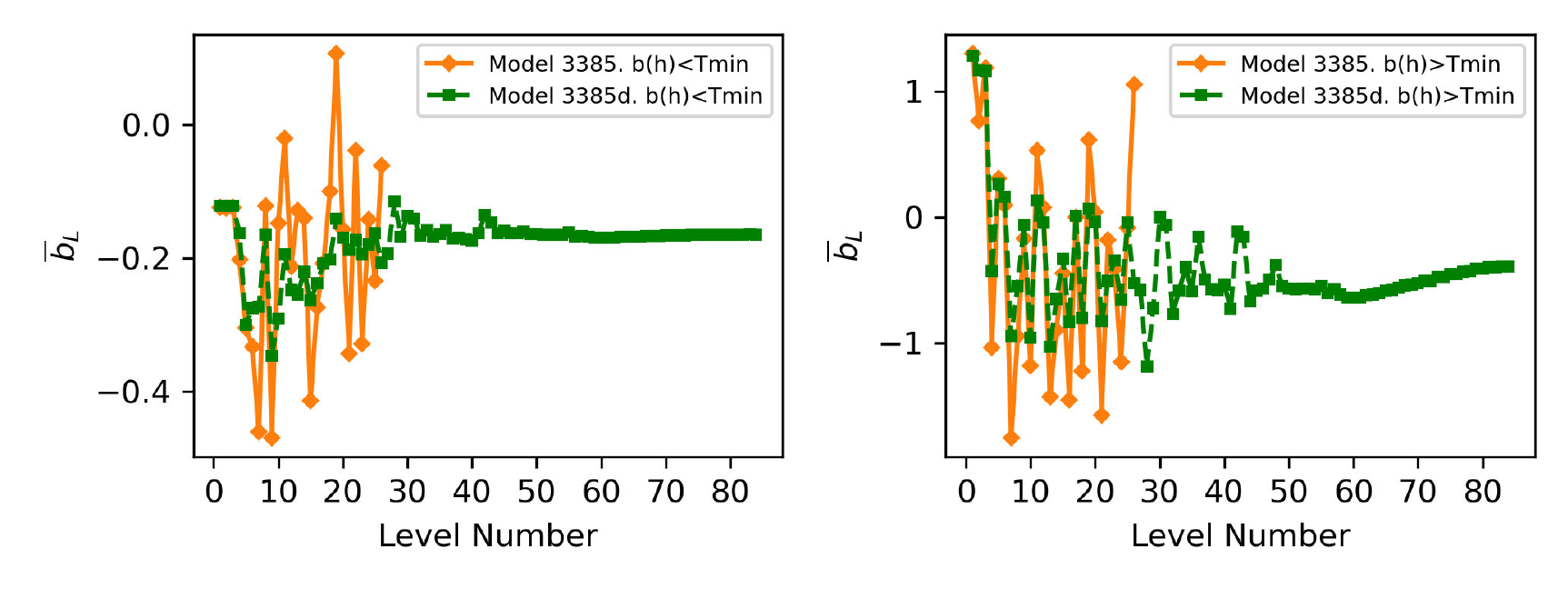}\vspace{0cm}
\includegraphics[width=0.8\textwidth,keepaspectratio]{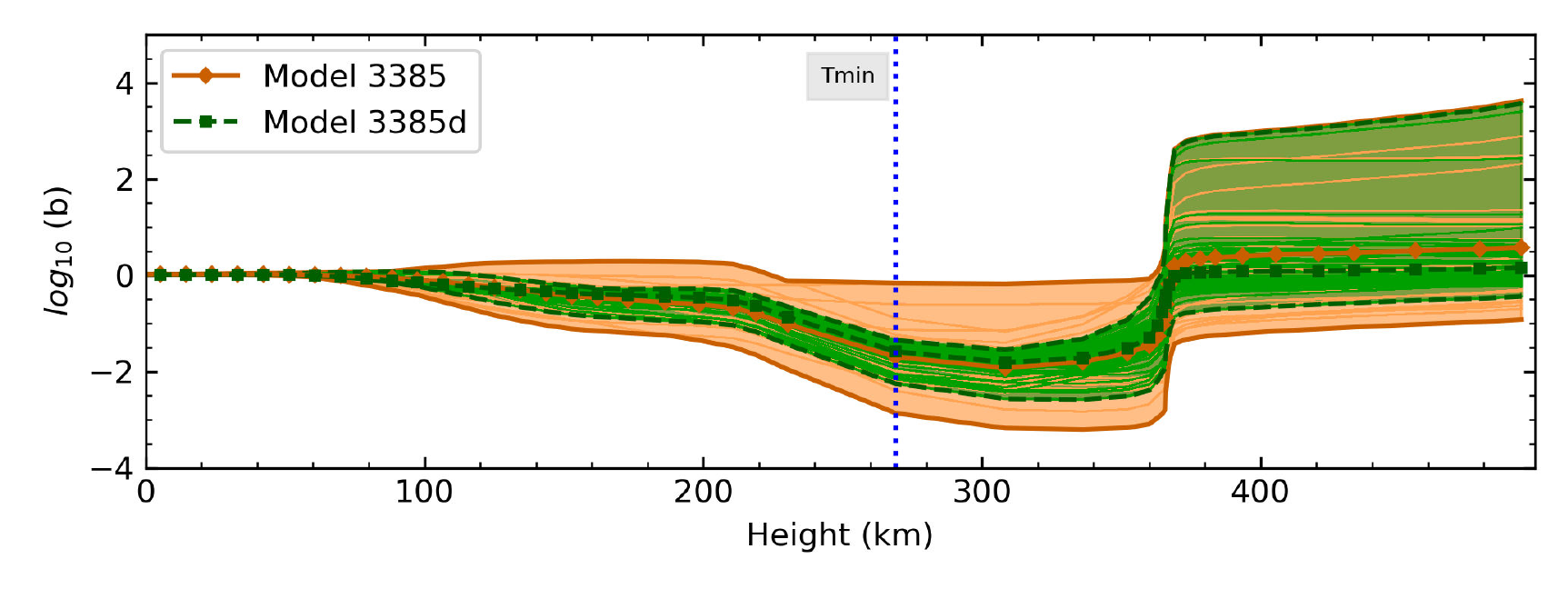}
\caption{Comparison of the distribution of \MgI in its energy levels (upper panel) and in the heights of the atmosphere (lower panel) between the 26-level base model (in orange) and the new 85-level model d (in green) for GJ 832. Details of subplots are the same as in Fig.~\ref{fig:b_Sun}.}
\label{fig:b_GJ832}
\end{figure*}

\begin{figure*}
\centering
\includegraphics[width=0.8\textwidth,keepaspectratio]{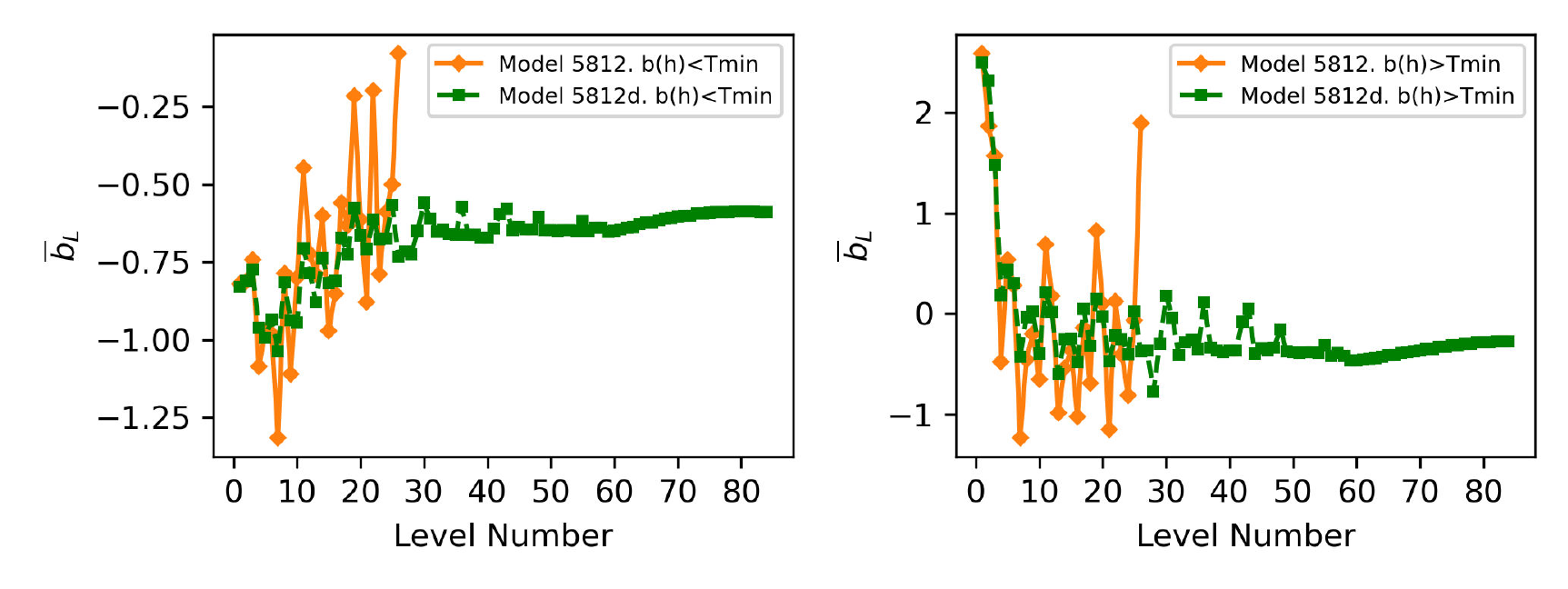}\vspace{0cm}
\includegraphics[width=0.8\textwidth,keepaspectratio]{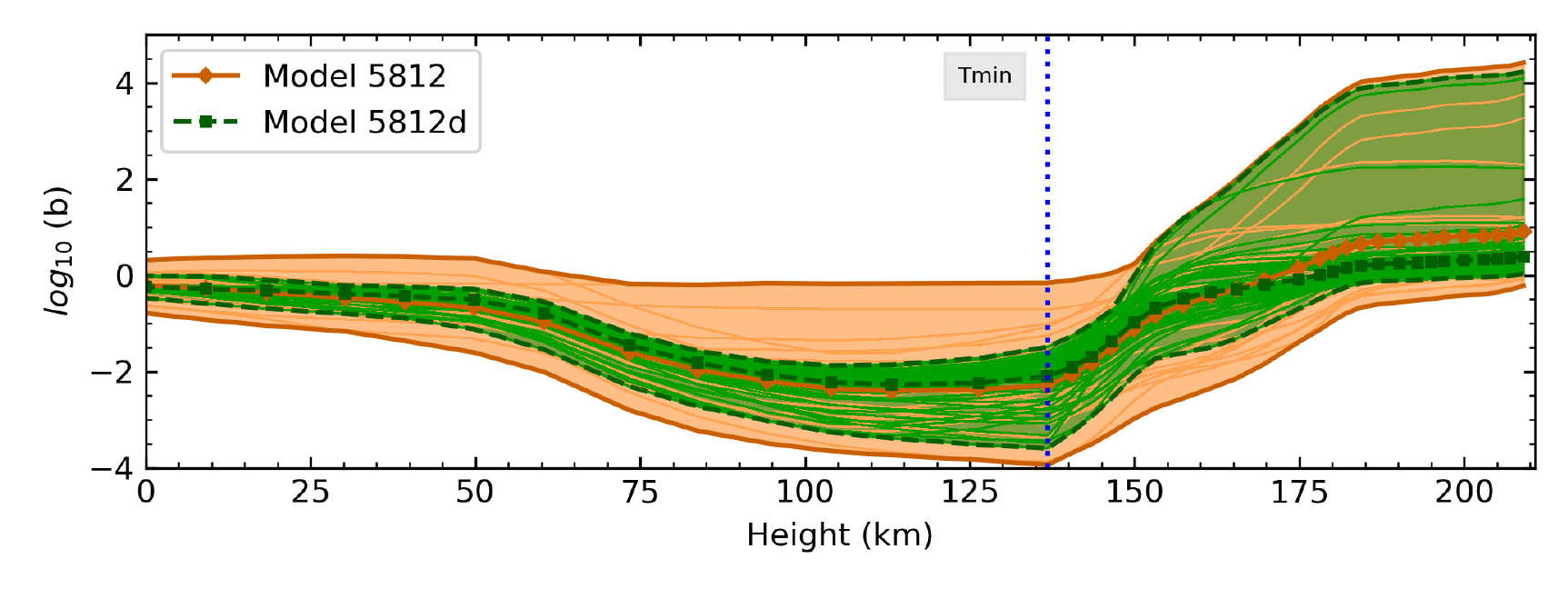}
\caption{Comparison of the distribution of \MgI in its energy levels (upper panel) and in the heights of the atmosphere (lower panel) between the 26-level base model (in orange) and the new 85-level model d (in green) for GJ 581. Details of subplots equal to Fig.~\ref{fig:b_Sun}.}
\label{fig:b_GJ581}
\end{figure*}

\textbf{\MgI distribution through energy levels and atmosphere heights}. We study the population changes through the LTE departure coefficient $b_L\,(h)$ \citep{rutten:2003} and
the \textit{height-averaged LTE departure coefficient} $\overline{b_L}$ defined in this work. We build this coefficient to get a general idea of the redistribution of the element of interest in its energy levels when the atomic model is being modified, but the atmospheric model remains the same. For a given level $L$ (of $N$ total levels), we calculate the average of $\log b_L\,(h)$ across the atmosphere to obtain an average departure coefficient value representative of the energy level: $\overline{b_L} = \sum_h \, [\log_{10} \, b_L(h)]\,/\,N$. 
The average is performed on the logarithms of the $b_L\,(h)$ so that the statistical weight of very small values (corresponding to the populations calculated in LTE, which are larger than those calculated in NLTE) 
is not affected in the final parameter. In addition, the averaging over the heights is split into two regions: from the beginning to the height where the temperature minimum ($T_{\text{min}}$) is located (the so-called photosphere) and from $T_{\text{min}}$ to the beginning of the transition region (defined as 0.95 $T_{TR}$, which is known as the chromosphere).

In Figs. \ref{fig:b_Sun}, \ref{fig:b_EERI}, \ref{fig:b_GJ832}, and \ref{fig:b_GJ581}, we show the height-averaged LTE departure coefficient $\overline{b_L}$ in the two regions described above as a function of the energy level number. 
These figures correspond to the Sun, Epsilon Eridani, GJ 832, and GJ 581, respectively.
The curves in the lower panels show the $\overline{b_L}$ departure coefficients for each model as a function of height. It is important to note that we have used the same scale for all of these subplots. In all cases, the base model (26 levels) and the new model d (85 levels) are shown.

For the four starts and both of the models, the \MgI LTE populations are generally 
larger than those in NLTE ($\log_{10}\,b<0$) in the photosphere. 
This behavior is expected due to the high density of \MgI (Fig.~\ref{fig:mg_distribution}) and electrons (right panel in Fig.~\ref{fig:T_ne_vs_press}), which favor a higher rate of e+\MgI collisions. 
The two-region separation described above is characterized by the height where the temperature minimum is reached and illustrated in the lower subplots of Figs.~\ref{fig:b_Sun}, \ref{fig:b_EERI}, \ref{fig:b_GJ832}, and \ref{fig:b_GJ581} with a vertical line. Starting at $T_{\text{min}}$, certain levels (especially the lower ones) 
begin to show a much larger $b_L\,(h)$ ($\log_{10}\,b > 0$, in this case) as it rises in the atmosphere toward the transition region.

Comparing the models to each other,  we see that the $b_L$ values in the four stars present lower dispersion in the energy levels (upper panels) in model d than in base. This effect can also be seen at different atmospheric heights (note the amplitude of each fringe in the lower panel). We attribute these results to a higher \MgI exchange between the energy levels, driven by the inclusion of a larger amount of radiative transitions (shown in Fig.~\ref{fig:histogram}) as well as the improved electron collisional data used. 
The new model's level populations are better coupled than those of the base model, leading to smaller population differences. The relatively large differences in $\overline{b_L}$ between the models for the same level are also strongly related to improvements in other atomic parameters, such as in the oscillator strength values, the photoionization rates, and the broadening parameters.

Comparing the models within the stars makes it  possible to obtain a first estimate of the result in the spectral lines. Noting the vertical scale between the top panels of Figs. \ref{fig:b_Sun}, \ref{fig:b_EERI}, \ref{fig:b_GJ832}, and \ref{fig:b_GJ581}, the $\overline{b_L}$ curves show that the population difference between models is, in most cases, smaller in the lower region of the atmosphere (top left panel) than in the upper region (top right panel); especially for the lower-lying levels ($\lesssim 5$). 

To analyze in detail a particular line in the star, one must study how the ratio of populations changes between the transition levels and the characteristics of the atmosphere in its formation region. A useful tool for this is the contribution function. 
The following Section will present details and exceptions in the spectral lines.

\subsection{Spectral lines} \label{subsec:lines}

This Section presents several line profiles calculated with both models, from FUV to MIR. We compare these values with observational data whenever possible, represented with dashed lines and circle symbols to illustrate the density of data points. We select the cases with the most significant change between models to exemplify the effects of the new atomic model of Mg I in the four stellar atmospheric models considered. Table~\ref{table:lines} presents the main characteristics of the spectral lines shown in this work.

\begin{table*}[ht]
\caption{Term-term line transitions shown in this work.}
\label{table:lines}
\centering
\begin{tabular}{ccccccc}
 \hline\hline
$\lambda_{vac}$\tablefootmark{(a)} & Transition & Transition Level Numbers & $\log \text{gf}$ \tablefootmark{(a)} & $\log \Gamma_4/N_e$\tablefootmark{(b)} & $\log \Gamma_6/N_H$\tablefootmark{(b)} & Fig.\\ ({\AA}) & \, & \textit{L(SL)--U(SU)} & \, & $(rad\,s^{-1}\,cm^3)$ & $(rad\,s^{-1}\,cm^3)$ & \,\\
\hline
 1\,747.8 & $3s^2\,^1S_0$--$6p\,^1P_1$ & 1(1)--25(1) & -2.04 & -4.52 & -7.07 & \ref{fig:1747}\\
 2\,026.5 & $3s^2\,^1S_0$--$4p\,^1P_1$ & 1(1)--9(1) & -0.95 & -5.62 & -7.43 & \ref{fig:2026}\\
 2\,780.6 & $3p\,^3P_{0,1,2}$--$3p^2\,^3P_{0,1,2}$ & 2(1,2,3)--28(1,2,3) & 0.75 & -5.98 & -7.70 & \ref{fig:sun}a\\
 2\,853.0 & $3s^2\,^1S_0$--$3p\,^1P_1$ & 1(1)--3(1) & 0.25 & -6.00 & -7.69 & \ref{fig:2853}\\
 6\,320.7 & $4s\,^3S_1$--$6p\,^3P_{0,1,2}$ & 4(1)--22(1,2,3) & -1.85 & -4.57 & -7.10 & \ref{fig:6320}, \ref{fig:sun}b\\ 
 8\,926.0 & $4s\,^1S_0$--$5p\,^1P_1$ & 5(1)--17(1) & -1.68 & -5.00 & -7.26 & \ref{fig:8926}\\ 
 10\,964.4 & $4p\,^3P_{0,1,2}$--$5d\,^3D_{1,2,3}$ & 7(1,2,3)--21(1,2,3) & 0.09 & -3.40 & -7.10 & \ref{fig:10964}, \ref{fig:sun}c\\ 
 33\,200.6 & $4d\,^3D_{1,2,3}$--$5f\,^3F_{2,3,4}$ & 13(1,2,3)--24(1,2,3) & 0.95 & -3.71 & -7.10 & \ref{fig:33200}, \ref{fig:sun}d\\ 
 71\,092.0 & $5f\,^1F_3$--$6g\,^1G_4$ & 23(1)--37(1) & 0.87 & -3.01 & -7.00 & \ref{fig:71092_71099}\\ 
 71\,097.4 & $5f\,^3F_{2,3,4}$--$6g\,^3G_{3,4,5}$ & 24(1,2,3)--38(1,2,3) & 1.35 & -3.01 & -7.00 & \ref{fig:71092_71099}\\ \hline\hline
\end{tabular}
\tablefoot{
\tablefoottext{a}{Extracted from the \textit{NIST} database (version 5.7.1).} \tablefoottext{b}{Broadening parameters from \cite{kurucz:1995}. $\Gamma_4$ and $\Gamma_6$ are given at 5\,000 K.}
}
\end{table*}

\begin{figure*}[t]
\centering
\includegraphics[height=0.45\textheight,keepaspectratio]{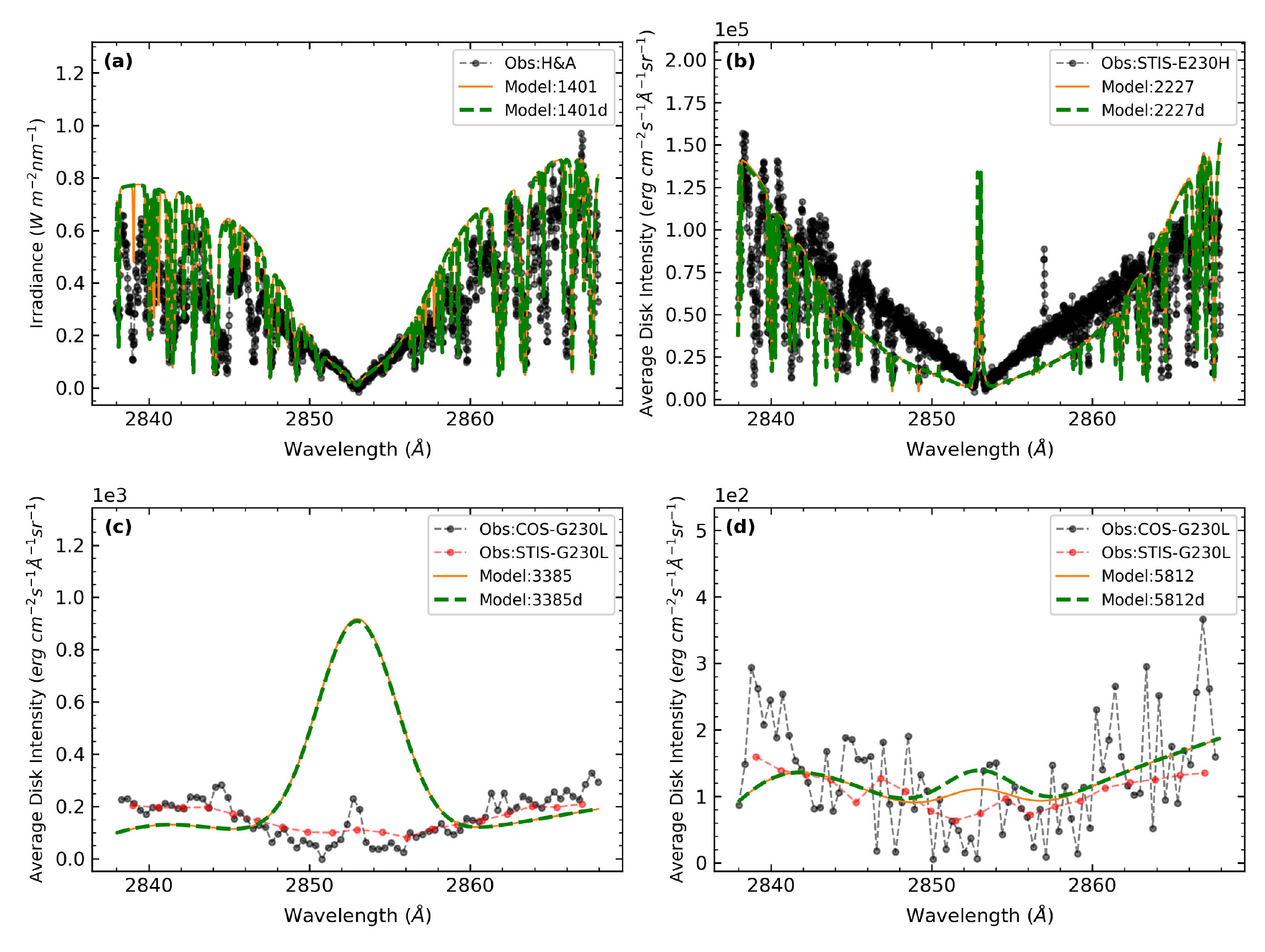}
\caption{Comparison of the spectral line 2853.0 {\AA} calculated with the base (orange solid line) and d (green dashed line) models over the observations (black or red dashed line with circle symbols),
for the Sun (a), Epsilon Eridani (b), GJ 832 (c), and GJ 581 (d).}
\label{fig:2853}
\end{figure*}

\begin{figure*}
\centering
\includegraphics[height=0.45\textheight,keepaspectratio]{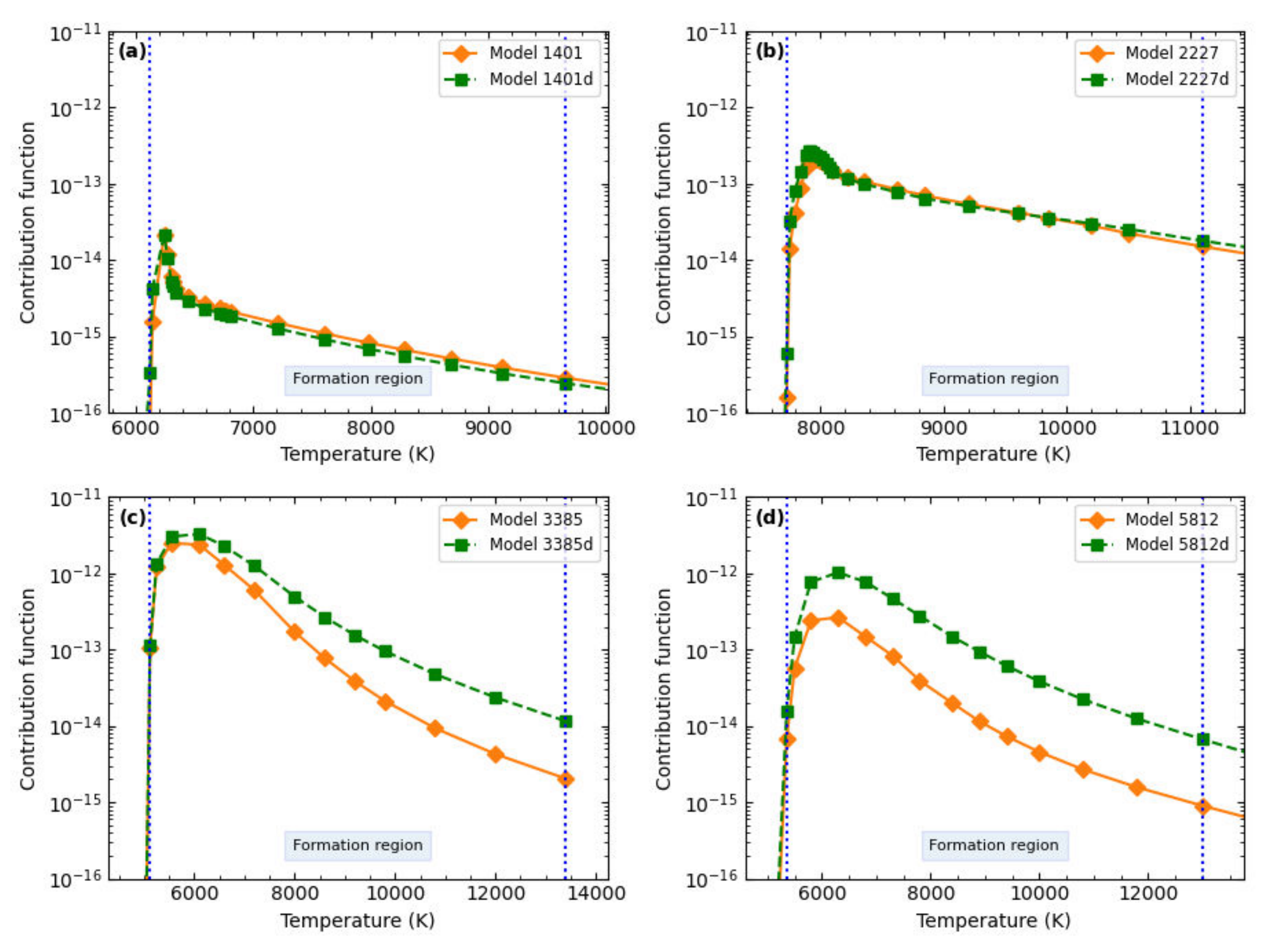}
\caption{Contribution function of the base (orange) and d (green) models for the Sun (a), Epsilon Eridani (b), GJ 832 (c) and GJ 581 (d) in the 2853.0 {\AA} line formation region.}
\label{fig:2853_fc}
\end{figure*}

\textbf{Analyzing the 2853.0 {\AA} line profile ($3s^2\,^1S_0$--$3p\,^1P_1$)}. Considering its formation height, the 2853.0 {\AA} line could be used as a diagnostic of the thermal structure of the stellar chromospheres. Although this transition is well described in the solar case by the base model, this is not true in stars cooler than the Sun. Using the base atomic model, neither the atmospheric model of \citet{fontenla:2016} nor those of \citet{tilipman:2021} have reached a correct fit for the observed line profiles of GJ 832 and GJ 581, taken with STIS-G230L (using $R=500$). A notable core emission is obtained in both cases, which is not present in the above-mentioned observations. Figure~\ref{fig:2853} compares the 2853.0 {\AA} lines obtained using the base and d models with observations for each star. For the dM stars (bottom panel), in addition to the STIS-G230L observations, we have added the COS-G230L observations (using $R=3000$). It is important to note that a central emission is found in the observations from STIS-E230H for Epsilon Eridani, and from COS-G230L for GJ 832. In the case of GJ 581, the low resolution of COS-G230L and the noisy signal mask the information, hindering the detection of a core emission. Furthermore, \citet{loyd:2016} recommend using COS as a reference to absolute flux levels, as the STIS instrument always measures lower values than COS, by factors of 1.1-2.4. With this in mind, it could be the case that STIS does not detect the peak correctly.

When comparing the models for the Sun and GJ 832, the differences between the synthetic lines calculated with the two models are negligible. In contrast, for Epsilon Eridani and GJ 581, the d model produces a larger intensity at the peaks and in the center. 
This result shows that the new model cannot solve the incorrect emission presented in the mentioned works. \citet{fontenla:2016} suggest the presence of the peak may be related to inaccuracies in the atomic collisional ionization rates and recombination data of \MgI. However, several tests performed by us on the atomic model parameters did not show an influence on the emission. In those tests, we were able to observe that the central emission could be avoided, without affecting the formation of other lines, when the upper-level population ($3s3p\,^1P$) is reduced by two orders of magnitude in the atmospheric region of the line formation. Assuming that the emission in the observations is not affected by any other factor, the evidence suggests an excess of \MgI at that level or the lack of some opacity medium in the cooler stars at that height.

The formation of a line can be studied by analyzing the population of the levels involved in the transition and the atmospheric conditions at the formation region. Figure~\ref{fig:2853_fc} shows the contribution function (or attenuated emissivity): \mbox{$f_c=(c^2/2h\nu) \, \varepsilon_\nu \, e^{-\tau_\nu}$} \citep{fontenla:2007} at the formation region of the line 2853.0 {\AA} for the  atmospheres of the different stars, calculated by the base and d models. The line formation mainly occurs around 6000 K for all the stars except for Epsilon Eridani, which presents a more significant contribution at 8000 K. Considering the left panel of Fig.~\ref{fig:T_ne_vs_press}, the main contribution to this line is in the chromospheric plateau for the Sun and Epsilon Eridani. For the cooler stars GJ 832 and GJ 581, the higher contribution is shifted to the right, closer to the transition region. Another notable aspect shown in Fig.~\ref{fig:2853_fc} is the order of magnitude of the $f_c$ values in each star (note that all the subplots present the same scale). The peak of the contribution function is higher in the atmosphere of the cooler stars, 
which becomes evident considering the local continuum level of each star. 

The contribution calculated with the two models is very similar for the Sun and Epsilon Eridani, but a considerable difference is obtained for GJ 832 and GJ 581. However, for GJ 832, the main difference is not found at the formation temperature of the line (the peak of the $f_c$), as occurs for GJ 581, which is in agreement with the synthetic line profiles produced by each model (panels c and d in Fig.~\ref{fig:2853}).

In the following, we present several lines already shown in Paper I for the Sun. To illustrate the impact of using different atomic models and the conditions of the atmospheric plasma of the star under consideration, we added the line profiles with the higher changes between the models and stars. These are an example of the importance of validating the same atomic model in stars of different spectral types.

\begin{figure*}
\centering
\includegraphics[height=0.45\textheight,keepaspectratio]{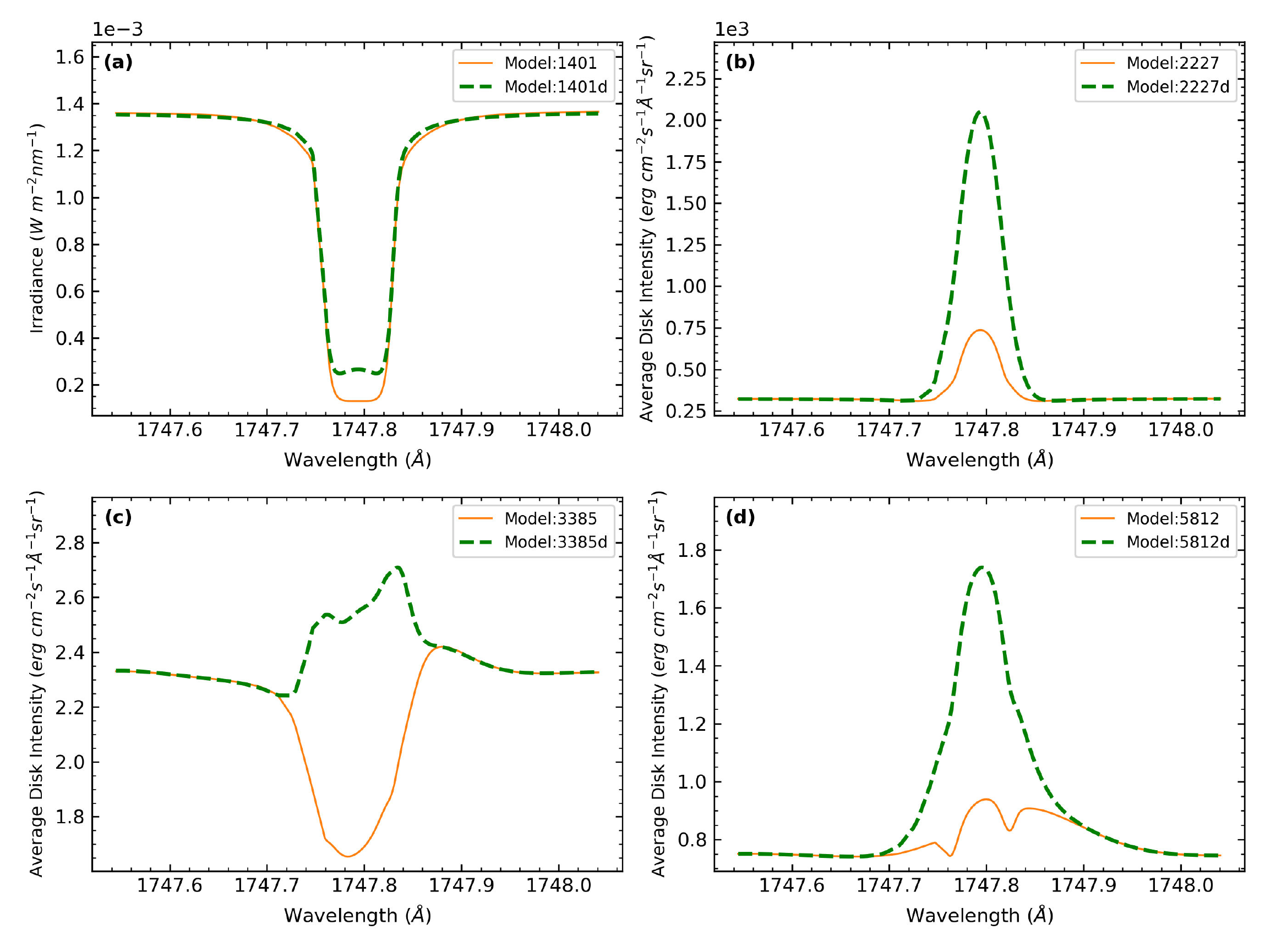}
\caption{Comparison of the spectral line 1747.7937 {\AA} calculated with the base (orange solid line) and d (green dashed line) models on the Sun (a), Epsilon Eridani (b), GJ 832 (c) and GJ 581 (d).}
\label{fig:1747}
\end{figure*}

\begin{figure*}
\centering
\includegraphics[height=0.45\textheight,keepaspectratio]{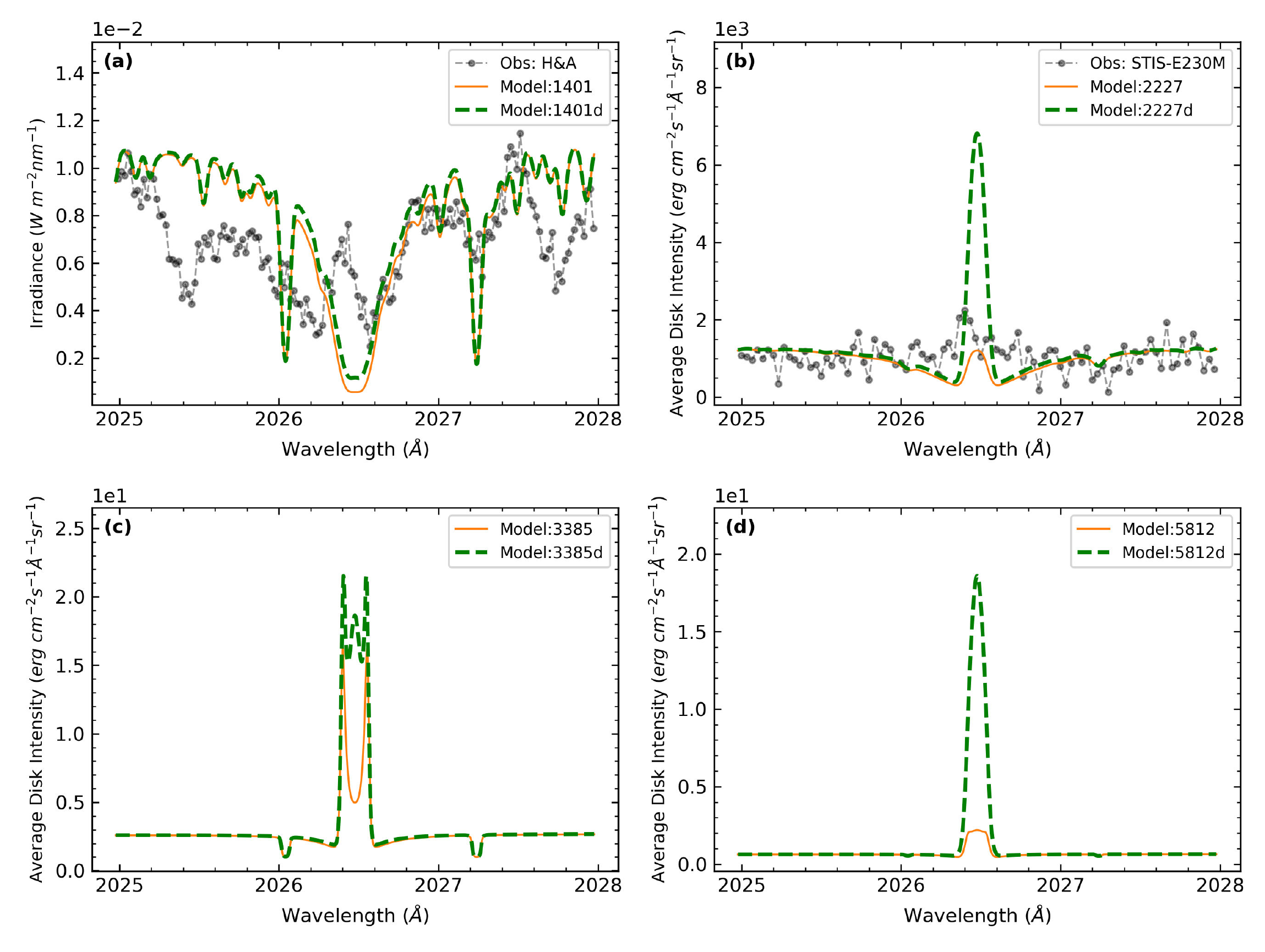}
\caption{Comparison of the spectral line 2026.4768 {\AA} calculated with the base (orange solid line) and d (green dashed line) models on the Sun (a), Epsilon Eridani (b), GJ 832 (c) and GJ 581 (d). Observations in the Sun and Epsilon Eridani are shown in black dashed lines with circle symbols.}
\label{fig:2026}
\end{figure*}

\textbf{FUV and NUV}. The most significant difference between the lines calculated with the base and d model is obtained in this spectral range. This result is consistent with the population study performed in Section \ref{subsec:d_populations}, that is the lines in this range are generally formed above the temperature minimum of each star. In addition to line 2853.0 {\AA}, the selected lines are 1747.8 {\AA} (Fig.~\ref{fig:1747}) and 2026.5 {\AA} (Fig.~\ref{fig:2026}), which present notable changes in gf of 17.2\% and -27.2\%, respectively. These lines are small and have a width of less than 0.5 {\AA}, so it was not possible to find observations where they were clearly noticeable. For line 2026.5 {\AA}, we used observations of the Sun and Epsilon Eridani. Despite that, we could observe different behaviors of the same spectral line between stars and models. In some cases, the effect is more significant in the center (even showing a reversal) and, in other cases, in the wings. Although a detailed analysis should be performed in each case, following the steps described in the example given above, the lines calculated by model d generally present a larger emission than the ones by the base model. This result is the final consequence of the multiple changes made to the base model (detailed in Section \ref{sec:new_model}), which had a greater impact on the lines formed in the chromosphere.

\begin{figure*}
\centering
\includegraphics[height=0.45\textheight,keepaspectratio]{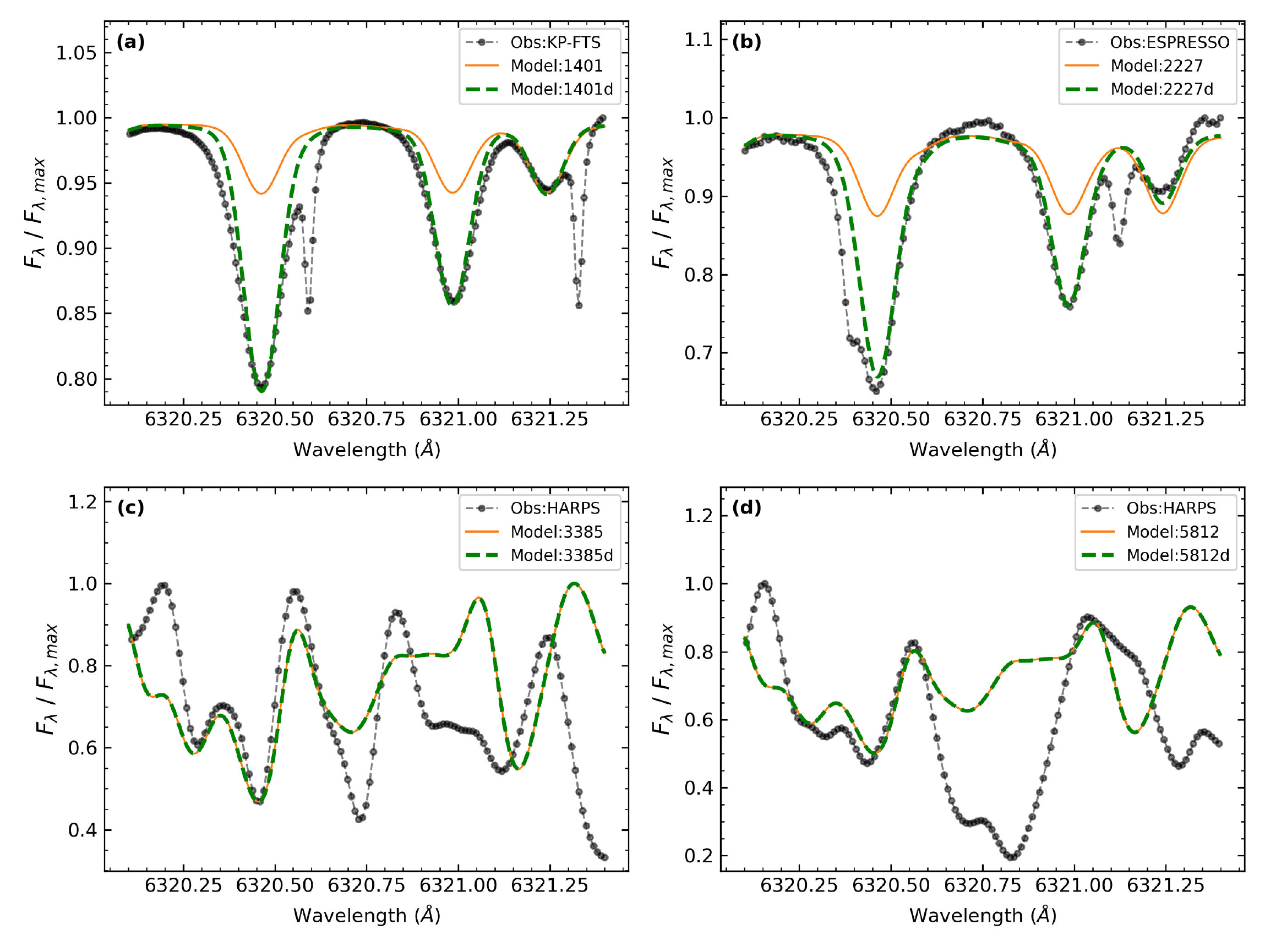}
\caption{Comparison of the spectral line 6320.7 {\AA} calculated with the base (orange solid line) and d (green dashed line) models on the Sun (a), Epsilon Eridani (b), GJ 832 (c) and GJ 581 (d). Observations are shown in black dashed lines with circle symbols.}
\label{fig:6320}
\end{figure*}

\textbf{VISIBLE}. Representing this range, Fig.~\ref{fig:6320} shows the 6320.7 {\AA} transition in each star. The differences obtained between the two atomic models are mainly due to the update of the radiative data. However, they are only noticeable in the Sun and Epsilon Eridani; in dM stars, this line is blended with molecular bands. In general, the variations between the lines produced by the base and d models are negligible in this spectral range. It is worth mentioning that the agreement with the observations is acceptable in all cases but remarkably good for the d model in the Sun and Epsilon Eridani.

\begin{figure*}[t]
\centering
\includegraphics[height=0.45\textheight,keepaspectratio]{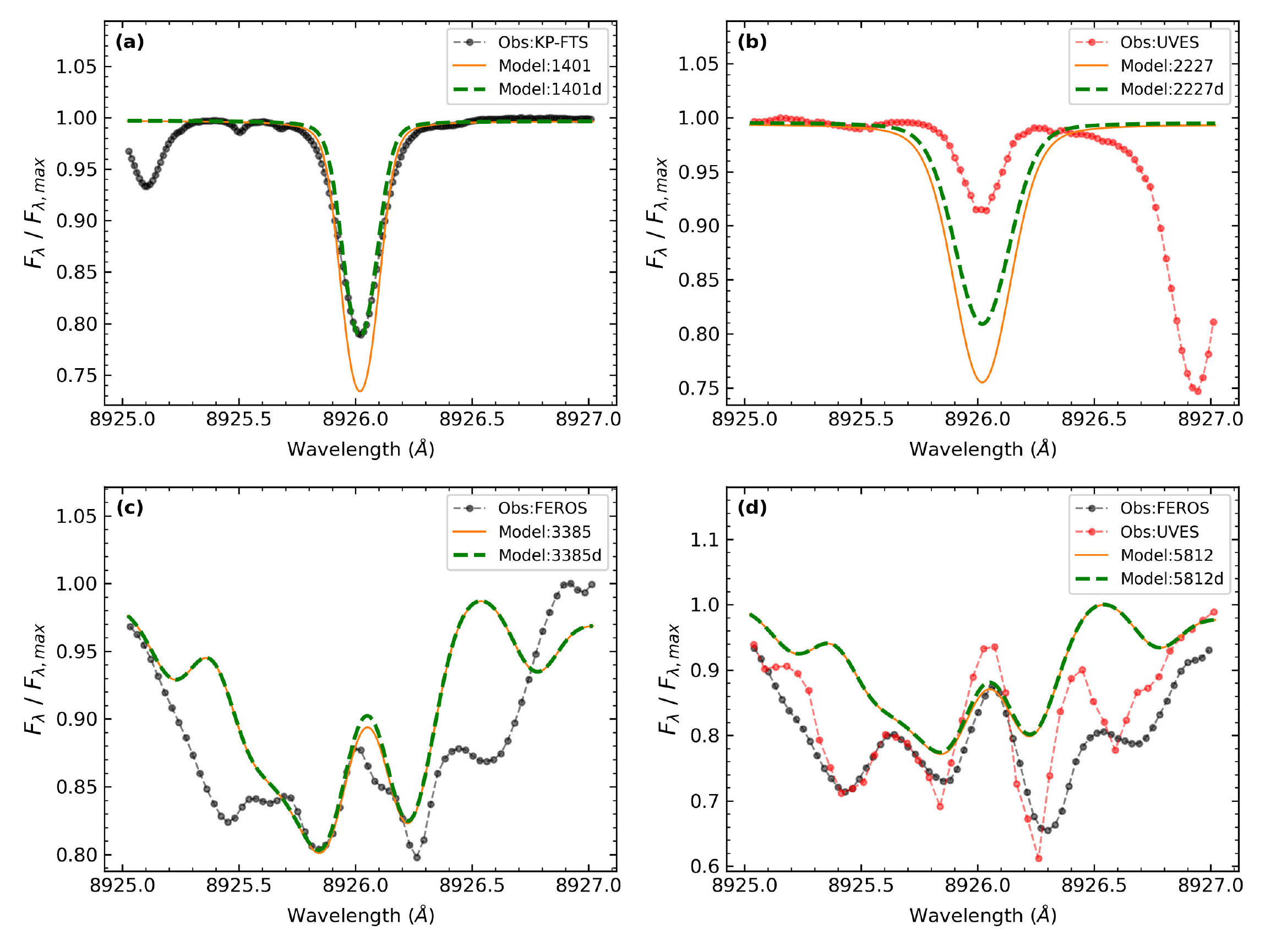}
\caption{Comparison of the spectral line 8926.0 {\AA} calculated with the base 
(orange solid line) and d (green dashed line) models on the Sun (a), Epsilon Eridani (b), GJ 832 (c), and GJ 581 (d). Observations are shown in black or red dashed lines with circle symbols.
}
\label{fig:8926}
\end{figure*}

\begin{figure*}
\centering
\includegraphics[height=0.45\textheight,keepaspectratio]{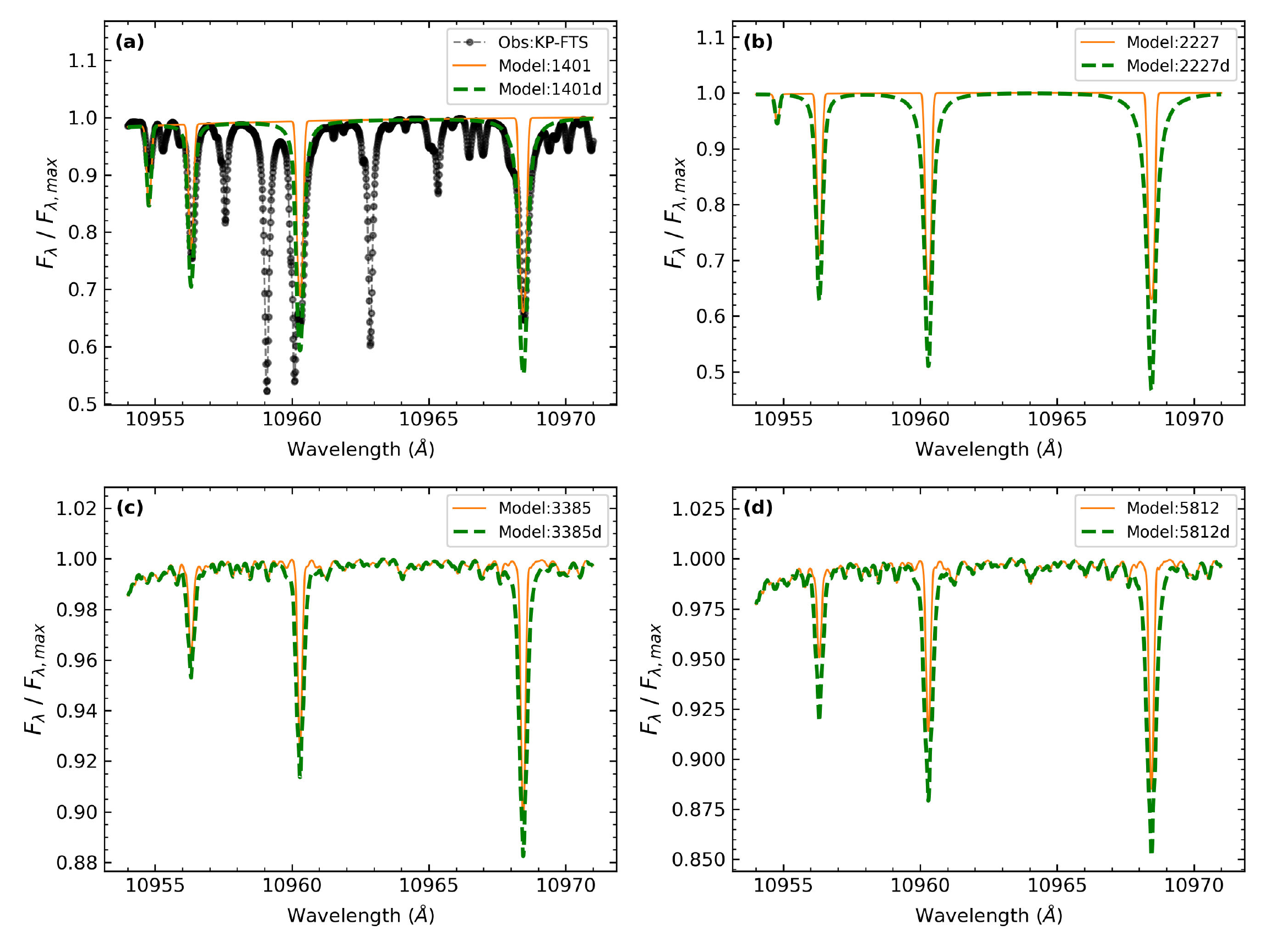}
\caption{Comparison of the spectral line 10\,964.4 {\AA} calculated with the base (orange solid line) and d (green dashed line) models on the Sun (a), Epsilon Eridani (b), GJ 832 (c) and GJ 581 (d). Observations in the Sun are shown in black dashed lines with circle symbols.
}
\label{fig:10964}
\end{figure*}

\textbf{NIR}. In this range, the lines calculated with both models are generally formed in the photosphere, well below the temperature minimum. However, significant changes between models, such as updating the gf values (Fig.~\ref{fig:8926}) or the addition of broadening data (Fig.~\ref{fig:10964}), become noticeable when the lower-lying energy levels of the transition are relatively high ($\gtrsim 5$, see the upper left panel in Figs \ref{fig:b_Sun}, \ref{fig:b_EERI}, \ref{fig:b_GJ832}, \ref{fig:b_GJ581}), depending on the star. In the above-given examples, model d can generate lower absorption in some cases and higher in others. The changes could even be much smaller between the different stars (top panel versus the bottom panel of Fig.~\ref{fig:8926}). The strong interdependence among the population densities across various energy levels poses a challenge in comprehending the causal and predictive effects of implementing the new d model. Nevertheless, considering the lines with the most significant differences between models and comparing them to the available observations, the d model reproduces the data better. An example of these results can be observed in the top panel of Fig.~\ref{fig:8926}, where the absorption produced by the new model matches almost exactly the data from KP-FTS in the Sun, and it is closer to the data from UVES for Epsilon Eridani than the line produced by the base model.

\begin{figure*}[t]
\centering
\includegraphics[height=0.45\textheight,keepaspectratio]{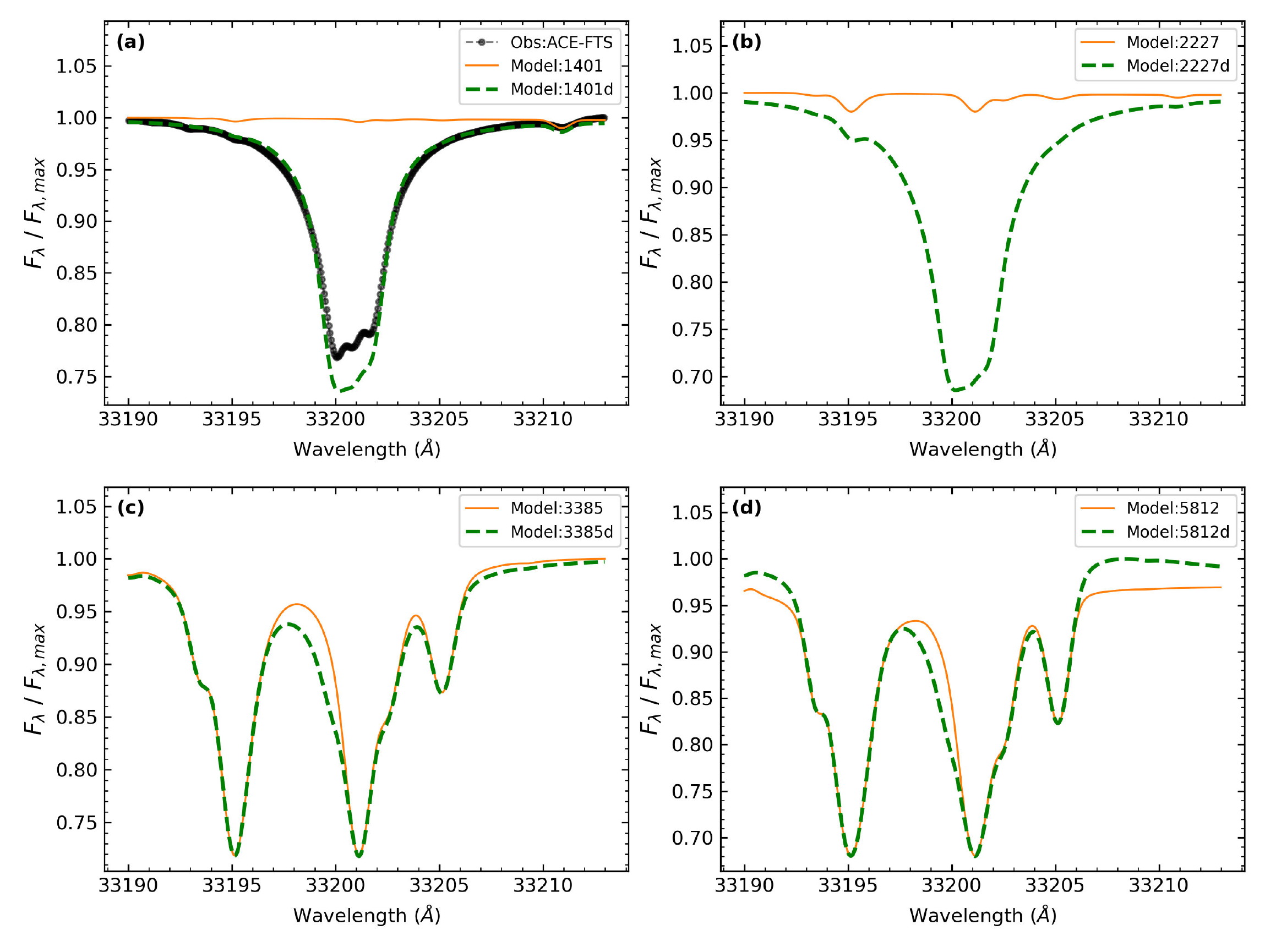}
\caption{Comparison of the spectral lines 33\,200.6 {\AA} calculated with the base (orange solid line) and d (green dashed line) models on the Sun (a), Epsilon Eridani (b), GJ 832 (c) and GJ 581 (d). Observations in the Sun are shown in black dashed lines with circle symbols.
}
\label{fig:33200}
\end{figure*}

\begin{figure*}[t]
\centering
\includegraphics[height=0.45\textheight,keepaspectratio]{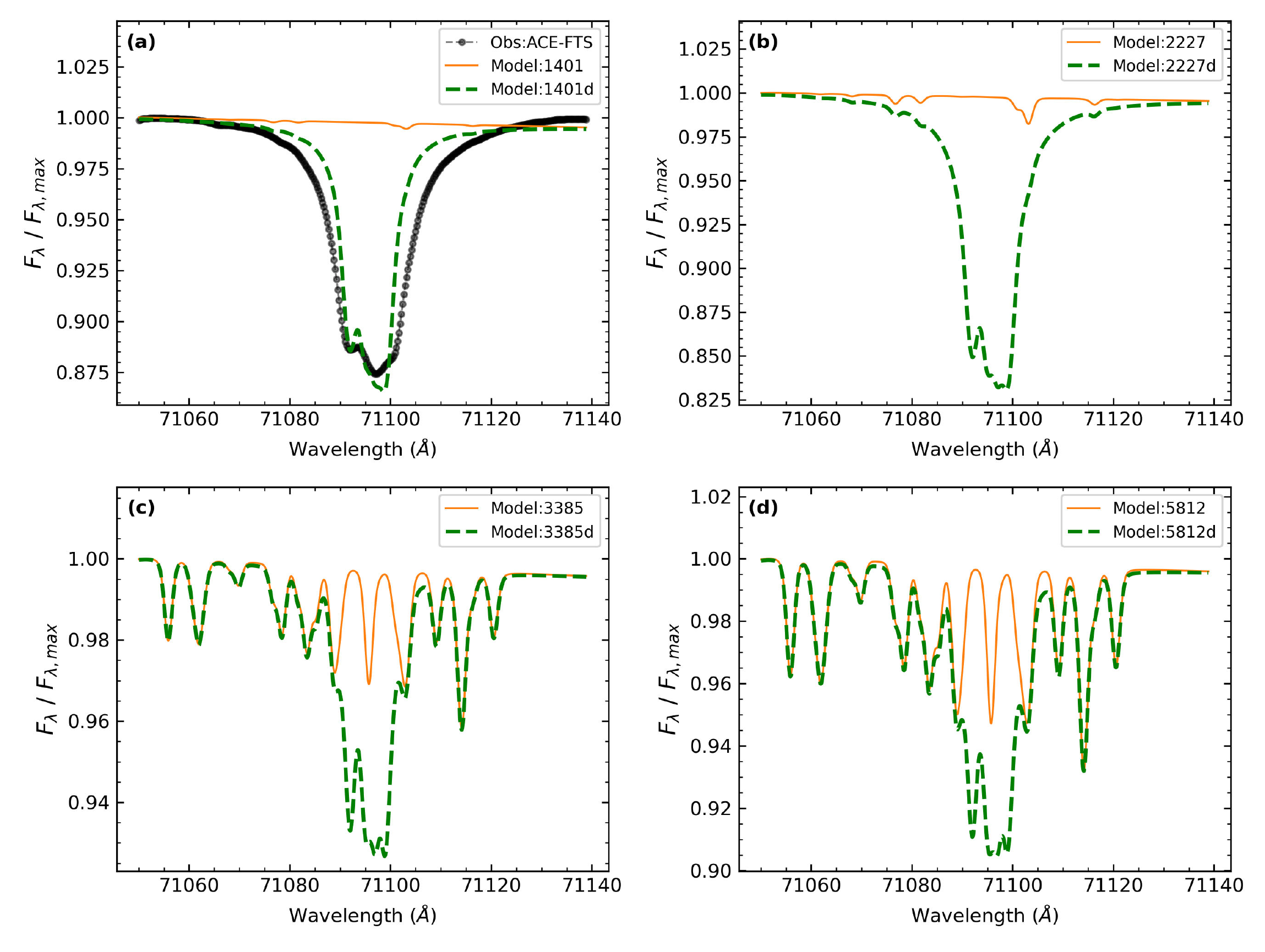}
\caption{Comparison of the spectral lines 71\,092 {\AA} with 71\,097.4 {\AA} calculated with the base (orange solid line) and d (green dashed line) models on the Sun (a), Epsilon Eridani (b), GJ 832 (c) and GJ 581 (d). Observations in the Sun are shown in black dashed lines with circle symbols.
}
\label{fig:71092_71099}
\end{figure*}

\textbf{MIR}. In this spectral region, we present three transitions, shown previously in Paper I for the Sun. They are formed well below the temperature minimum, but unlike the lines in the NIR, most occur between high energy levels and, therefore, cannot be reproduced by the base model. As seen in some of the previous examples, lines 33\,199~{\AA} (Fig.~\ref{fig:33200}), 71\,092~{\AA}, and 71\,097.4~{\AA} (Fig.~\ref{fig:71092_71099}) can be compared in two well-distinguishable groups: the Sun and Epsilon Eridani, and the dM stars. Within each group, the d model produces similar lines, likely due to the characteristic continuum level of each spectral type in this region. The only observations available in this range are those of ACT-FTS transmittance for the Sun. It can be observed that, in both figures, the d model presents a very good match with these measurements. Future observations in this range will allow us to extend our study to other stars and, therefore, improve our atmospheric and atomic models.


\section{Conclusions} \label{sec:conclusions}

Improving the atomic model of \MgI for the NLTE populations calculation using the SRPM code libraries produced a solar spectrum in good agreement with the different observations from NUV to MIR, as shown in Paper I (c model). This improvement consisted of updating parameters such as Einstein coefficients, broadening parameters (Stark and van der Waals, mainly), and updating and including photoionization data, energy levels, and spectral lines. The model was extended from 26 to 85 levels, with superlevels from index 55 onward, which allowed 127 new spectral lines to be reproduced, extending the maximum wavelength previously set at 17 000 {\AA} to 71 500 {\AA}. 
In addition, the electron impact excitation effective collision strengths ($\Upsilon_{ij}$) for \MgI, which have thus far used the semiempirical formulas of Seaton and van Regemorter (SEA\&VRM), were replaced with quantum-mechanical ones.
Namely, data from \citet{barklem:2017} calculated via the CCC method were used for the first 25 levels. 
Multiconfiguration Breit-Pauli DW calculations from Paper I were used for the first time in this species for levels from 26 to 54. These levels mainly participate in the formation of lines in the IR range.
The $\Upsilon_{ij}$ data for transitions involving superlevels from level 55 onward, were completed with SEA\&VRM values. Finally, for \MgII, the above parameters were also improved, and the number of levels was extended from 14 to 47. These changes allowed us to reproduce 729 new spectral lines, covering a wide range from 850 {\AA} to 630 um. 

For the present work, we started from the c model and further improved the \MgI atomic model. The new model (d) differs from model c in the following ways: regarding the data for the e+\MgI collisions, the DW calculation was extended to the levels where SEA\&VRM formulas were previously used. Hence, the new model is formed by CCC data in the first 25 levels and our DW calculations from level 26 onward, including the superlevels. Regarding the radiative data, we included 5674 theoretical transitions (3001 term to term), which were also calculated by us.

On the Sun, the new \MgI model showed minimal differences relative to the c model in the spectral lines that belong to the FUV, NUV, and visible regions, with more significant differences in the NIR and MIR ranges. Considering that the solar atmospheric model of \citet{fontenla:2015} was built with an outdated \MgI atomic model, a new atmospheric model should be built to fit the observations up to the MIR region correctly.

We extended the study of the \MgI model made for the Sun in Paper I to three stars cooler than the Sun: Epsilon Eridani, GJ 832, and GJ 581. For this purpose, we studied and compared the populations and spectra between the original, called base, and d models. Comparing the NLTE population obtained with both models, the following was noticed:
\begin{enumerate}
    \item For the first ionization states (\MgI, \MgII, and \MgIII) and the molecules containing Mg, it was observed that \MgII predominates in the stellar atmospheres of the Sun and Epsilon Eridani, with more than 95\%; whereas, in GJ 832 and GJ 581, the presence of \MgI is more significant, with more than 72\%. In the latter, it was also observed that the amount of Mg forming molecules is more than two orders of magnitude higher than in the Sun and Epsilon Eridani. On the other hand, when analyzing the total population changes between the base and d models, the latter showed an increase in \MgI greater than 2\% in the Sun and Epsilon Eridani, which is correlated with a decrease in \MgII and \MgIII with respect to their total densities. This suggests a migration of ionized magnesium toward \MgI, probably through recombination processes. 
    This process is favored in the d model by a \MgI electronic structure closer to the continuum level, which is not featured in the base model. This argument is reinforced by comparing the same models in cool stars where, for both GJ 832 and GJ 581, migration occurs from \MgI to \MgII and \MgIII.
    \item For the \MgI energy levels, and for different heights in the atmosphere (from the photosphere to the base of the transition region), we used the LTE departure coefficient ($b_L$) and the LTE departure coefficient averaged over heights ($\overline{b_L}$ ), constructed by us. From analyzing both parameters, we could see that:
    \begin{enumerate}
        \item The LTE departure is smaller for heights below  $T_{min}$ than for above this value, in agreement with what is expected.
        \item The d model showed less dispersion than the base model in its departure coefficients, both among the different energy levels and at different atmospheric heights. This result shows a better-coupled model population.
        \item The largest differences in $b_L$ between the models considered occurred after the temperature minimum and at levels approximately from five onward.
    \end{enumerate}
\end{enumerate}

In addition to the populations, the \MgI spectral lines produced by the models were also studied. The impact of a change in the atomic data on a given spectral line depends on the population change in the involved levels and the characteristics of the atmosphere in the region where the line is formed. As a result, for the same population variation in the levels, the formation of some lines may be more affected than others, and even the same line in a different star. The most noteworthy aspects of model d on the spectral lines are as follows:
\begin{enumerate}
    \item In the FUV and NUV regions, the most pronounced differences with the base model are observed. The behavior of the same spectral line can vary between stars and \MgI atomic models. In some cases, the effect is more pronounced in the center of the line (even showing a reversal), while in others, it is more pronounced in the wings. Overall, it can be seen that model d produces more emissions than the base model.
    
    At the 2853.0 {\AA} line, none of the changes made to the atomic data resolve the incorrectly calculated core emission reported by \citet{fontenla:2016} and \citet{tilipman:2021}. However, it is important to note that observations of GJ 832 obtained by COS-G230L, which date from the same time as the observations used in constructing the models in both works, show central emission, although with much less intensity than calculated by SSRPM. \citet{loyd:2016} state that there was a discrepancy in the flux measurements by COS and STIS, where the cause of the systematic low-flux observed by STIS could be due to an incorrect alignment of the spectrograph slit on the target. Furthermore, observations obtained by STIS-E230H of Epsilon Eridani show that this line has a central emission in this star. This could mean that the line depends, to some degree, on the activity level of the star being studied. If we assume that the emission in the observations is not being affected by another factor, the intensity of the emission could be due to an excess of \MgI in that level or the lack of some NUV opacity medium. To avoid the central line emission, maintaining the formation of other lines unchanged, the population of the upper level ($3s3p\,^1P$) in the atmospheric formation region should be reduced by two orders of magnitude.
    \item In the visible, the differences between models were generally negligible. Although there are exceptions, the \MgI lines in this range are usually formed in the photosphere, and they are formed by transitions between relatively low levels; consequently, the changes made to the base model do not affect them significantly. The observations found for this range allowed us to further verify the accuracy of our atomic and atmospheric models.
    \item In the IR, it was observed that model d can generate lower absorption in some cases and higher in others, and even the change can be much lower between different stars. In the MIR, we observed that the lines in the Sun and Epsilon Eridani are similar to each other, as are those in the dM stars. This is possibly due to the similarity at each star's continuum level. We did not find observations for the stars; however, for the Sun, the d model can reproduce the observed spectral lines very well.
    
    The IR lines strongly depend on collisions, so in order to calculate and use them as indicators (of activity, abundance, etc.), it is essential to have reliable atomic data.
\end{enumerate}

Finally, it is important to note that when using the new atomic model of \MgI to calculate the atmospheric model for the Sun and the other stars, it could produce spectral lines that differ from the observed spectrum. In this case, the atmospheric models should be corrected to fit the line formation provided by the new atomic model over the entire spectral range covered by it.


\begin{acknowledgements} \label{sec:acknow}
We sincerely thank Dr. Jeffrey Linsky for his detailed revision of our manuscript since it has helped us improve our work.
This work has used the \textit{VALD} database, operated at Uppsala University, the Institute of Astronomy RAS in Moscow, and the University of Vienna.
We acknowledge to 1995 Atomic Line Data (R.L. Kurucz and B. Bell) Kurucz CD-ROM No. 23. Cambridge, Mass.: Smithsonian Astrophysical Observatory.
This research was supported by grants PICT 2018-2895 and PICT 2019-4342 from the Agencia Nacional de Promoción Científica y Tecnológica (MINCyT, Argentina).
\end{acknowledgements}

\bibliographystyle{aa_url} 
\bibliography{JIPrefs} 

\end{document}